\documentclass[fleqn,usenatbib]{rasti}

\usepackage{newtxtext,newtxmath}

\usepackage[T1]{fontenc}

\DeclareRobustCommand{\VAN}[3]{#2}
\let\VANthebibliography\thebibliography
\def\thebibliography{\DeclareRobustCommand{\VAN}[3]{##3}\VANthebibliography}

\usepackage{graphicx}	
\usepackage{amsmath}	
\usepackage{aas_macros}
\usepackage{svg}

\newcommand{\msun}{\mathcal{M}_{\sun}}
\newcommand{\orcid}[1]{\href{https://orcid.org/#1}{\includesvg[width=10pt]{orcid}}}

\title[WD Photometric Toolkit]{WDPhotTools -- A White Dwarf Photometric Toolkit in Python}

\author[M. C. Lam et al.]{
M. C. Lam$^{1, 2\,\orcid{0000-0002-9347-2298}}$\thanks{Contact e-mail: \href{mailto:lam@mail.tau.ac.il}{lam@mail.tau.ac.il}},
K. W. Yuen$^{3}$,
M. J. Green$^{1\,\orcid{0000-0002-0948-4801}}$, and
W. Li$^{1\,\orcid{0000-0002-0096-3523}}$\\
$^{1}$School of Physics and Astronomy, Tel Aviv University, Tel Aviv, Israel 69978\\
$^{2}$Astrophysics Research Institute, Liverpool John Moores University, IC2, LSP, 146 Brownlow Hill, Liverpool L3 5RF, UK\\
$^{3}$Department of Ophthalmology \& Visual Sciences, Faculty of Medicine, The Chinese University of Hong Kong
}

\date{Accepted 2022. Received 2022; in original form 2022}

\pubyear{2022}

\begin{document}
\label{firstpage}
\pagerange{\pageref{firstpage}--\pageref{lastpage}}
\maketitle

\begin{abstract}
From data collection to photometric fitting and analysis of white dwarfs
to generating a white dwarf luminosity function requires numerous Astrophysical,
Mathematical and Computational domain knowledge. The steep learning curve makes
it difficult to enter the field, and often individuals have to reinvent the wheel
to perform identical data reduction and analysis tasks. We have gathered a wide
range of publicly available white dwarf cooling models and synthetic photometry
to provide a toolkit that allows (1)~visualisation of various models,
(2)~photometric fitting of a white dwarf with or without distance and reddening,
and (3)~the computing of white dwarf luminosity functions with a choice of
initial mass function, main sequence evolution model, star formation history,
initial-final mass relation, and white dwarf cooling model. We have recomputed
and compared the effective temperature of the white dwarfs from the Gaia EDR3
white dwarf catalogue. The two independent works show excellent agreement in
the temperature solutions.
\end{abstract}



\section{Introduction}
White dwarfs~(WDs) are the final stage of stellar evolution of main
sequence~(MS) stars with zero-age MS~(ZAMS) mass less than
$\sim$\,$8\msun$~\citep{2013sse..book.....K}. Since this
mass range encompasses the vast majority of stars in the Galaxy, these
degenerate remnants are the most common final product of stellar evolution,
thus providing a good sample to study the history of stellar evolution and star
formation at the earliest time of the Galaxy. In this state,
there is little nuclear burning to replenish the energy they radiate away. As a
consequence, the luminosity and temperature decrease monotonically with time.
The electron degenerate nature means that a WD with a typical mass of
$0.6\mathcal{M}_{\sun}$ has a similar size to the Earth, giving rise to their
high densities, low luminosities, and large surface gravities.

The use of the white dwarf luminosity function~(WDLF) as a cosmochronometer was
first introduced by \citet{1959ApJ...129..243S}. Given the finite age of the
Galaxy, there is a minimum temperature below which no white dwarfs can reach in
a limited cooling time. This limit translates to an abrupt downturn in the WDLF
at faint magnitudes. Evidence of such behaviour was observed by
\citet{1979ApJ...233..226L}, however, it was not clear at the time whether it
was due to incompleteness in the observations or to some defect in the
theory~(e.g.,~\citealp{1984ApJ...282..615I}). A decade later,
\citet{1987ApJ...315L..77W} gathered concrete evidence for the downturn and
estimated the age\footnote{``Age'' refers to the total time since the oldest
WD progenitor arrived at the zero-age main sequence.} of the disc to be
$9.3 \pm 2.0$\,Gyr~(see also \citealt{1988ApJ...332..891L}). While most studies
focused on the combined Galactic thin and thick discs~\citep{1989LNP...328...15L,
1992ApJ...386..539W, 1995LNP...443...24O, 1998ApJ...497..294L, 1999MNRAS.306..736K,
2012ApJS..199...29G, 2021A&A...649A...6G}, some worked with the combined discs
and the stellar halo~\citep{2006AJ....131..571H, 2017AJ....153...10M,
2019MNRAS.482..715L}. There was one attempt in disentangling the thin disc,
thick disc and the stellar halo component WDLFs \citep{2011MNRAS.417...93R}.
 
Most WDs have similar broadband colour to main sequence stars,
so they cannot be identified using photometry alone. They are
found from UV-excess, large proper motion and/or parallax. Because of the
strongly peaked surface gravity distribution of WDs, photometric fitting for
their intrinsic properties is possible by assuming a surface gravity. WDs
fitted in such a way are statistically useful provided that the
sample is not
strongly biased. This is demonstrated in various studies comparing photometric
and spectroscopic solutions to calibrate the atmosphere
model~\citep{2019ApJ...871..169G, 2019ApJ...882..106G}, as well as from the
agreeing shapes of the WDLFs from spectroscopic and photometric samples. The
Gaia satellite provides parallactic measurements for over a billion point
sources~\citep{2021A&A...649A...1G, 2021AJ....161..147B} of which $359\,000$
are high confidence WD candidates~\citep[][hereafter, GF21]{2021MNRAS.508.3877G}.
The availability of parallaxes allows much more accurate fitting, particularly,
without knowing the surface gravity for the photometric sample. This has
completely  revolutionized the field of WD sciences. In the forthcoming decade,
the Simonyi Survey Telescope at the Vera C. Rubin Observatory will continue to
discover more WDs at fainter magnitudes, but only accompanied by proper
motion measurement at best at the faintest magnitudes.
Furthermore, at those magnitudes, it is infeasible to collect spectra for most
of them and thus studies will mostly rely on photometric methods.

In this generation of user-side astronomical data handling and
analysis, as well as in the computing courses for scientists, \textsc{Python} is
among the most popular languages due to its ease to use with a
relatively shallow learning curve, readable syntax and a simple way to `glue'
different pieces of software together. Its flexibility to serve as a scripting
and an object-oriented language makes it useful in many use cases: demonstrating
with visual tools with little overhead, prototyping, and web-serving, and it can
be compiled if wanted. This broad range of functionality and high-level usage
make it relatively inefficient. However, Python is an excellent choice of
language to build wrappers over highly efficient and well-established codes. In
fact, some of the most used packages,
\textsc{scipy}~\citep{2020NatMe..17..261V} and
\textsc{numpy}~\citep{2020Natur.585..357H}, are written in
\textsc{Fortran} and \textsc{C}, respectively.
\textsc{WDPhotTools} depends heavily on these two highly optimised and matured
packages to deliver the desired efficiency.

This paper is organised as follows: in Section~\textsection2, we describe
the software structure and development process. Section~\textsection3 describes
the model formatter. Section~\textsection4 covers the photometric fitting
procedures and the overview of the model used. It also explains the
construction of a theoretical luminosity function including some descriptions
of the models available. Section~\textsection5 goes through the other tools
that are essential to the software. We describe some auxiliary tools that is
also available with this toolkit in Section~\textsection6, and we summarise this
work in Section~\textsection7.

Throughout this work, we use $m$ to denote apparent magnitude, $M$ for
absolute magnitude, curly $\mathcal{M}$ to denote mass, where $\mathcal{M}_i$
for the (initial) ZAMS mass, and $\mathcal{M}_f$ for the (final) WD mass.

\section{Software and Development Organisation}
The goal of this work is to ease researchers the effort in reinventing the
wheel for trivial, repetitive but essential first steps in many
aspects of data analysis for WD science. \textsc{WDPhotTools} allows a simple
setup to compare results from different choices of models. We note that this work is only suitable for the analysis of the
most common types of WDs. It is not advisable to perform analysis on any WDs
that are suspected to have evolved with companion, or with atypical core
and/or atmospheric composition.
There is already a spectroscopic version for a similar purpose, the
\textsc{wdtools} \footnote{\url{https://github.com/vedantchandra/wdtools}}~
\citep{2020MNRAS.497.2688C}. While we are drawing
an analogy between the two pieces of software, we would like to emphasise that their
implementation starts from a lower level that directly makes use of the
output products of the stellar evolution models. The synthetic
photometry part of the \textsc{WDPhotTools} begins with pre-computed
tables in which the filter profiles have already been convolved with the model
spectra, with the normalisation (absolute magnitude) of the photometry
determined based on the mass-radius relation from their simulations. Future
development of \textsc{WDPhotTools} will move away from this~(see
Section~\textsection7).

\textsc{WDPhotTools} can generate colour-colour diagrams,
colour-magnitude diagrams in various photometric systems,
plot cooling profiles from different models, and compute
theoretical WDLF based on the built-in or supplied models.
The core parts of this work are three-fold: (1)~the tailored-formatters that
handle the output models from various works in the format as they were
generated and downloaded. This allows the software to be updated to handle
future releases of the models easily. They are written as
two base classes: \texttt{AtmosphereModelReader} and
\texttt{CoolingModelReader}; (2)~photometric fitter
(\texttt{WDFitter} class), that solves for
the WD parameters based on the photometry, with or without distance and
reddening; and (3)~to generate white dwarf luminosity function in
bolometric magnitudes or in any of the photometric systems available from the
atmosphere models (with the \texttt{WDLF} class).

We host our source code on Github\footnote{\url{https://github.com/cylammarco/WDPhotTools}},
which provides version control and other utilities to facilitate the
development; the numbered releases are also deposited on Zenodo to allow
referencing of a specific version\footnote{\url{https://doi.org/10.5281/zenodo.6595029}}.
It offers issue and bug tracking, high-level project management,
automation with Github Actions upon each commit for:

\begin{enumerate}
    \item Continuous Integration (CI) to install the software in Linux, Mac
    and Windows system, and then perform unit tests with \textsc{pytest}
    (Krekel et al. 2004);
    \item generating test coverage report with Coveralls\footnote{\url{
    https://coveralls.io/github/cylammarco/WDPhotTools}}
    which identifies
    lines in the script that are missed by the tests;
    \item Continuous Deployment (CD) through \textsc{PyPI} that allows
    immediate availability of the latest software release; and
    \item generating application programming interface~(API) documentation
    powered by \textsc{Sphinx} hosted on
    Read the Docs\footnote{\url{https://wdphottools.readthedocs.io}}.
\end{enumerate}

The following is not serving as an API, it is describing some of the key
arguments and parameters, as well as the usage that may seem convoluted but
is necessary to keep the structure simple without providing
multiple methods to perform similar tasks with small adjustments. The
model-readers are designed to read the files as they are downloaded from the
source, such that the future updates of the models provided by the same research
group should be a straightforward process in extending the model list.
The materials that have gone into the classes and modules are
described in \textsection3 to \textsection6 alongside with the usage of those
classes and modules. We summarise this work in \textsection7.

\section{Formatters}
There are two classes of model readers, one is for reading the synthetic
photometry computed from the Montreal DA/DB atmosphere models; the other one is
a formatter that handled the different formats in the output files from
different providers.

\subsection*{Synthetic Photometry}
The publicly available Montreal models\footnote{\url{https://www.astro.umontreal.ca/~bergeron/CoolingModels/}},
provide synthetic photometry of WDs over a smooth grid of wide ranges of temperature
and surface gravity. The tables are provided with bolometric and absolute
magnitudes on various photometric systems in pure hydrogen~(DA) and pure-helium~(DB)
atmospheres and in the range of $\log(g)=7.0 - 9.0$, as well as
the effective temperature and total cooling time~(though this grid is sparse
compared to their complementary cooling model grids). The
photometric systems include the Johnson-Kron-Cousins~($U, B, V, R\,\&\,I$),
Two Micron All Sky Survey~(2MASS) $J, H\,\&\,K_{s}$, Mauna Kea Observatory~(MKO)
$Y, J, H\,\&\,K$, Wide-field Infrared Survey Explorer~(WISE) $W1, W2, W3\,\&\,W4$,
Spitzer Space Telescope Infrared Array Camera~(IRAC)
$[3.6], [4.5], [5.8]\,\&\,[8.0] \mu m$, Sloan Digital Sky Survey~(SDSS)
$u, g, r, i\,\&\,z$, Panoramic Survey Telescope and Rapid Response
System~(Pan-STARRS\,1) $g, r, i, z\,\&\,y$,
Gaia $G, G_{\mathrm{BP}}\,\&\,G_{\mathrm{RP}}$, and Galaxy Evolution
Explorer~(GALEX) $FUV$ and $NUV$.

The stellar masses and cooling ages are based on the latest generation of
evolution sequences~\citep{2020ApJ...901...93B} with the choices of
thick~($q_H \equiv \frac{\mathcal{M}_H}{\mathcal{M}*} = 10^{-4}$) and thin~($q_H = 10^{-10}$)
hydrogen layers that are referred to as the pure-hydrogen and
pure-helium model atmospheres, respectively. Details of the colour calculations
are described in \citet{1995PASP..107.1047B} and \citet{2006AJ....132.1221H}.
The DA grid covers a temperature range $T_{\mathrm{eff}} = 2\,500 - 150\,000$\,K
while the DB grid covers $T_{\mathrm{eff}} = 3\,250 - 150\,000$\,K. Both model
are computed with surface gravities $\log(g) = 7.0 - 9.0$~\citep{
2018ApJ...863..184B, 2020ApJ...901...93B, 2011ApJ...730..128T,
2011ApJ...737...28B, 2006ApJ...651L.137K}.

\subsection{Class: \texttt{AtmosphereModelReader}}
\begin{figure}
    \centering
    \includegraphics[width=\columnwidth]{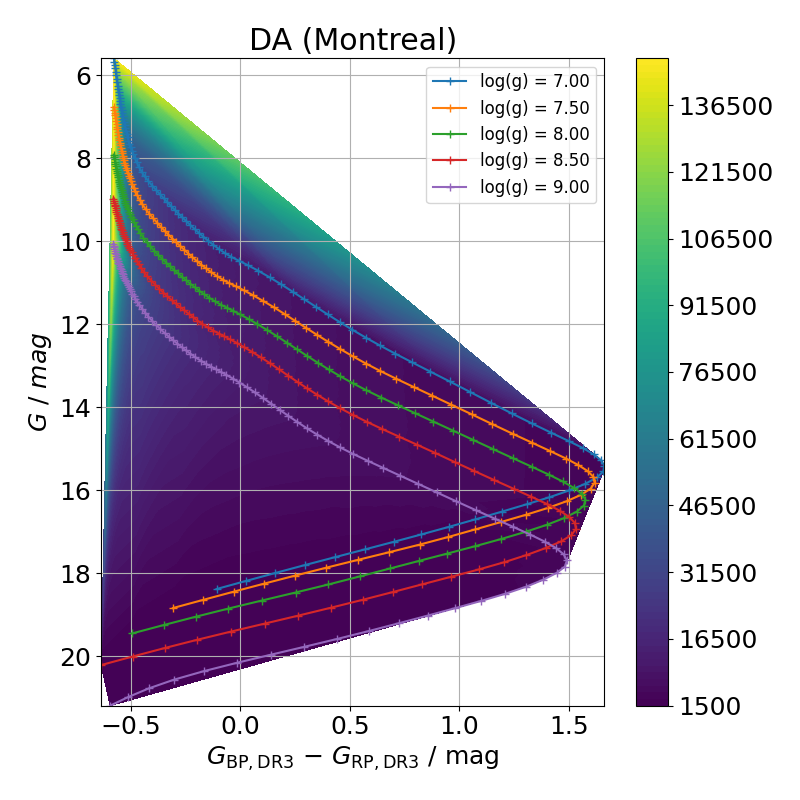}
    \caption{Default figures showing the cooling tracks at different surface
    gravities in the Gaia $G$, $G_{\mathrm{BP}}$ and $G_{\mathrm{RP}}$ filters.
    The colours scale shows the effective temperature.}
    \label{fig:cooling_tracks_default}
\end{figure}

This class takes care of the formatting of the models. As compared to the
cooling model reader, this is a simple reader because only the DA and DB
atmosphere types in the same output format are available for a wide range of
parameters and in various commonly used broadband filters. After initialisation,
the method \texttt{list\_atmosphere\_parameters()} can be used to retrieve the
parameters available for interpolation with \texttt{interp\_am()}. By default, it
is initialised as a DA atmosphere interpolator providing the interpolated $G$
band magnitudes in the Gaia DR3 photometric system~(dependent), parameterised
in \texttt{logg} and \texttt{Mbol}~(independent). We name the arguments dependent
and independent using the common terminology in mathematical modelling and
experimental science: the variables defining the axes of the interpolation are
the \textit{independent} variables, and the variable being interpolated over
is the \textit{dependent} variable. The number of independent variables can be
one or two. In the former, the argument \texttt{logg} has to be provided which
is defaulted to $8.0$. There are two choices of interpolator:
\texttt{scipy.interpolate.CloughTocher2DInterpolator()} and 
\texttt{scipy.interpolate.RBFInterpolator()}. The former is faster but it is less
stable at the grid boundaries, and it also does not
natively support extrapolation. The
latter is slower\footnote{This method is available since \textsc{scipy}
version 1.7, however, we strongly recommend using version 1.9 or above where
there is a significant performance gain in the \texttt{RBFInterpolator}.} but it is more
stable against irregularly sampled data as well as at the boundaries. It also
supports extrapolation natively. The choice depends on the usage. The
default is choosing \texttt{CloughTocher2DInterpolator} because
it would be usable with older distribution of \textsc{scipy}. Fine control
of the interpolator can be performed using keyword arguments. The cooling
tracks on a colour-magnitude diagram are shown for the Montreal DA model
in Figure~\ref{fig:cooling_tracks_default}.

\subsection{Class: \texttt{CoolingModelReader}}
The reader for the 24 cooling models from sixteen pieces of work
is more complex because the outputs from different models are
organised differently. These are all handled in the background and users only
need to choose among the models. Similar to the \texttt{AtmosphereModelReader},
after initialisation, the method \texttt{list\_cooling\_model()} can be used to
retrieve the available models and then the parameters available for
interpolation from those models can be retrieved using
\texttt{list\_cooling\_parameters()}. With the chosen model name and parameter
names, an interpolator can be built using the method \texttt{interp\_cm()}, which
will use the appropriate formatter to read the model files already contained in
the package. In this class, the same two interpolators,
\texttt{scipy.interpolator.CloughTocher2DInterpolator()} and
\texttt{scipy.interpolate.RBFInterpolator()}, are available for use. See
Table~\ref{tab:cooling_models} for the complete listing of the
available models.

\section{Photometric fitting}
Photometric fitting with synthetic broadband photometry has been the common
practice in getting the first guess of the atmospheric properties of white
dwarf candidates, or any unknown astronomical source. Photometry from multiple
filters and parallactic measurements can improve on the fitting accuracy.
It may seem to become obsolete in the era of Gaia as most white
dwarfs have measured parallaxes. However, the faint targets with respect to any
survey are always going to remain a challenge as the uncertainties in all the
observables are going to be significant. Hence, photometric fitting will
continue to serve as a useful method in the future. Once VRO starts producing
WD candidates at faint magnitudes, they would be far beyond the capabilities
of most spectrographs. The few spectrographs mounted on 10\,m$+$ telescopes are a
rare resource for follow-up and it would be not possible to obtain spectra
to confirm a sizeable sample of WD candidates. We will have to rely mostly
on photometric method again, and without any parallactic measurements.

\subsection{Class: \texttt{WDFitter}}
\label{sec:wdfitter}
This class inherits from the \texttt{AtmosphereModelReader}. For basic usage,
users are only required to provide the atmospheric type(s), the magnitudes and
uncertainties in the respective filters. For more advanced usage, other
fine-tuning can be performed by providing distance, extinction, independent
variables\footnote{Fitting using \texttt{Mbol} and \texttt{Teff} can give slightly
different results due to interpolation.}, choice of interpolator and their
fine-tuning parameters, minimisers and their fine-tuning parameters,
or solution refinement within a bounding box~(only if \textsc{emcee} was used).
A few arguments are worth extra attention because of the similar terms used
by different methods. We are only pointing out the potential confusion with
the arguments, the detailed usage should be referred to the API document
itself:
\begin{enumerate}
    \item The argument \texttt{extinction\_convolved} refers to the use of
    tabulated values from convolving the filter response function with DA
    spectra to compute the extinction corrected for the spectral shape
    (see also Section~\textsection4.3 and Figure~\ref{fig:interstellar_reddening})
    When it is set to \texttt{False}, it interpolates the tabulated values
    from the Appendix of \citet{2011ApJ...737..103S}.
    \item \texttt{kind} is the kind of interpolation for the extinction.
    \item \texttt{logg} is used \textit{if only} one independent variable is
    provided.
    \item \texttt{method} sets the choice from the 3 minimizers.
    \item There are 5 keyword arguments that can be used to
    perform fine control over the minimizer and
    interpolator: \texttt{kwargs\_for\_RBF}, \texttt{kwargs\_for\_CT},
    \texttt{kwargs\_for\_minimize}, \texttt{kwargs\_for\_least\_squares} and
    \texttt{kwargs\_for\_emcee}.
\end{enumerate}

\begin{figure}
    \centering
    \includegraphics[width=\columnwidth]{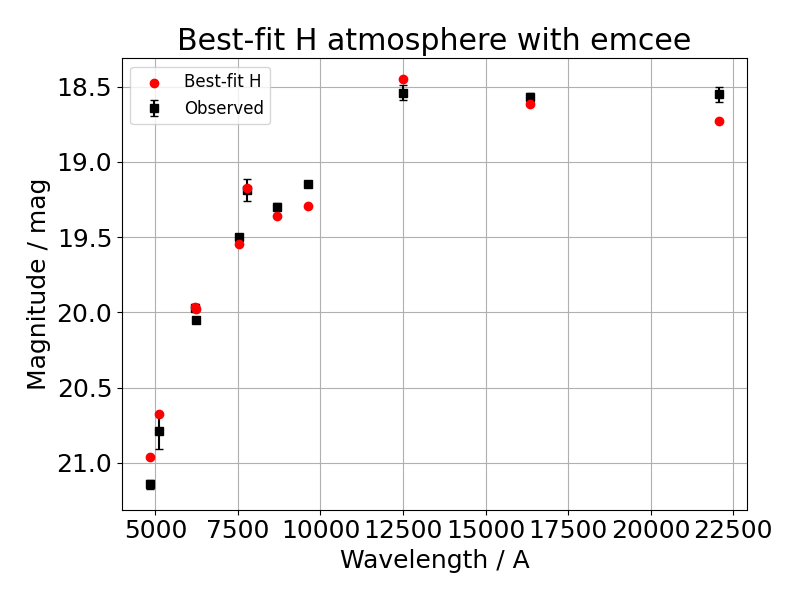}
    \includegraphics[width=\columnwidth]{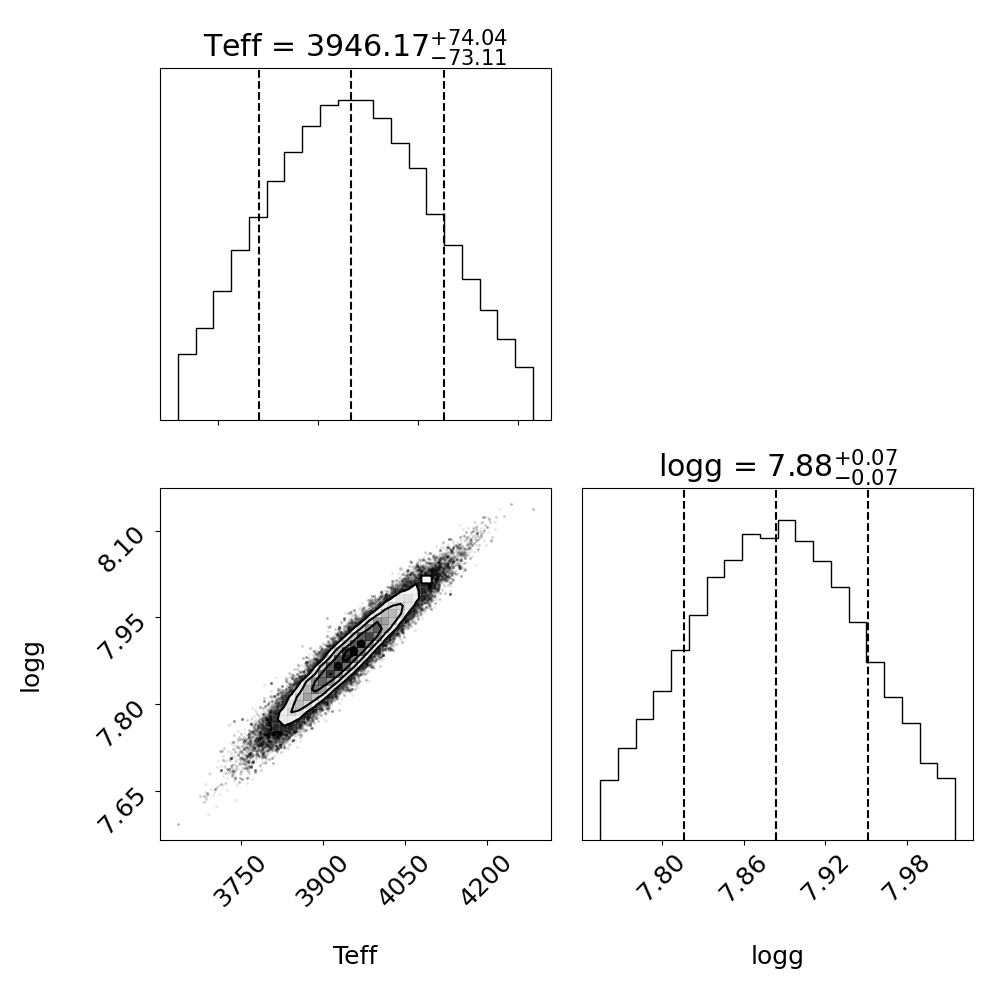}
    \caption{Best fit DA solutions found using MCMC with \textsc{emcee} and the associated corner plot of the
    sampling for PSO\,J1801+6254~\citep[][they found a spectroscopic of T$_\mathrm{eff}=3550$\,K]{2020MNRAS.493.6001L}.}
    \label{fig:best_fit}
\end{figure}

This class also comes with two diagnostic plots. The first one,
\texttt{show\_best\_fit()} overplots the best fit photometry corrected for the
(fitted) distance (and reddening) with the observed
data. The second one, \texttt{show\_corner\_plot()},
shows the probability distribution functions from the MCMC
sampling~\ref{fig:best_fit}.

Photometric fitting can be performed by minimizing the $\chi^2$ with any method
supported by \texttt{scipy.optimize.minimize()},
\texttt{scipy.optimize.least\_squares()} or with a Markov chain Monte
Carlo method powered by \texttt{emcee}~\citep{2013PASP..125..306F} with the
option to refine the solution with \texttt{minimize()} at the end using the
50$^{\mathrm{th}}$-percentile as the initial guess and bounding the fit
within N-sigmas uncertainty limit (user-defined). The
uncertainties are not computed when using the {\texttt{minimize()}} method, while
the latter two do estimate the uncertainty as part of the minimisation
procedure. At high temperatures ($T_{\mathrm{eff}} \geq 11\,000$\,K), the
differences in mass inferred from pure H or pure He models are of the order
of $\sim$\,$0.1\,\msun$. However, the differences between the pure He and
mixed H/He solutions in the same temperature range are completely negligible.
At lower temperatures, the atmospheric compositions have much stronger effects.
When fitted with pure He models, the average mass of a population decreases;
the average mass decreases further when fitted with the mixed H/He models. The
above-average-mass WDs in the temperature range of $6\,000\,-10\,000$\,K
have an average mass above $0.7\,\msun$ when fitted with pure hydrogen model,
but this drops to $\sim$\,$0.6\,\msun$ when fitted with mixed
H/He models~\citep{2019ApJ...876...67B}. Below $\sim$\,$6\,000$\,K, optical
spectra are not particularly useful in distinguishing the atmosphere
type, because hydrogen absorption lines \textit{start} to disappear, while the
helium lines disappear
as early as at $\sim$\,$9\,000$\,K~\citep{2018ApJ...857...56R}. However, the
broad spectral features, namely the collisionally induced absorption, in the
infrared can distinguish the atmosphere type~(see Figure~12 of
\citealt{2017ApJ...848...36B}). Therefore, readers are reminded to pay
particular care when you arrive at a low temperature solution with only
optical data.

With $n$ observables, it is only possible to fit for $n-1$ variables. If there
are only two photometric bands, it is only possible to fit
for one parameter if the \textit{distance is unknown}. By combining 2-band
photometry \textit{with a distance}, there is enough degree of freedom to fit two
parameters.

This does not apply to fitting with or without interstellar reddening because
it is merely a one-to-one lookup value, which itself is
only a function of distance. When studying a population of WDs that are fitted
for the distance, we should compare the photometric distance to the distances
measured from other means, e.g. astrometry, spectroscopic distance etc. in order
to assess the bias in the fit~(e.g.\ Section~\textsection3.2.2
in \citealt{2011MNRAS.417...93R}).

The effective temperature (and absolute bolometric magnitude) and the solid
angle of a source can be obtained when the distance is known, through the
relation $\Omega = \pi R^2 / D^2$. In the era of
Gaia~\citep{2021A&A...649A...1G}, most WD candidates have
their parallaxes measured, and hence the distances
derived~\citep{2021AJ....161..147B}. This allows much more reliable photometric
fitting with broadband photometry alone, as compared to fitting with an assumed
surface gravity~(see below). By using an evolutionary model of WDs, the radius
and the effective temperature can be used to obtain the mass~(and surface
gravity). We do not provide a means to use an alternative
mass-radius relation of white dwarfs. The fitting is only performed using the
relation already adopted in the tables of synthetic photometry.

Even though this is known as the photometric fitting, and values are almost
exclusively reported in magnitude, the measurement of the apparent brightness
is ultimately performed in electron counts. With good calibration and detector
linearity behaviour, it is sufficiently accurate to treat the uncertainties
at the flux level. The uncertainty in flux measured has a Gaussian-like error
that is symmetrical to both over- and under-estimation of the flux. However, because magnitudes are in the logarithmic space of the flux, the
uncertainty in magnitude is no longer symmetrical about the ``central'' value.
Thus fitting directly with magnitude weighted by the magnitude uncertainty
leads to a bias: solutions are biased to be over-luminous. For example, at
$\pm0.1$\,mag, the flux ratios are $1.0965$ and $0.9120$, respectively,
corresponding the $(-)8.8\%$ and $(+)9.65\%$ change in the flux in the faint
and bright end. This only narrows down to $(-)0.917\%$ and $(+)0.925\%$ at
$\pm0.01\,$mag level\footnote{A negative change in the magnitude corresponds
to an increase in flux.}. This effect becomes visually obvious if one is to
perform an MCMC sampling and inspect the probability distribution functions.
For this reason, in \textsc{WDPhotTools}, we fit in flux-space where the linear
and symmetrical approximation of the uncertainty distribution is more
realistic.

For comparison against other work on model fitting, it is very important to
state the minimisation/likelihood function. If the functions are different to
begin with, it is almost certainly impossible for the recomputed data to
have a perfect agreement with the data set that is compared against. Secondly, the
different choices (or even versions) of interpolation and minimisation methods
will also prevent repeatability. Thus, we detail the likelihood functions
and the choice of the interpolators as follows.

\subsection{Mathematical Construction}
We provide three means of fitting. The first two are both finding the minimum
$\chi^2$, with different minimizers. In the following equations, we use $m$ and
$f$ to denote apparent magnitude and flux; and $M$ and $F$ to denote absolute
magnitude and flux. The function to be minimized in these two cases is the
square of the weighted difference between model and observation.
\begin{equation}
    \label{eq:lsq}
    \left(\dfrac{F_{\mathrm{model}, i} - F_{i}}{\sigma_{F, i}}\right)^{2}
\end{equation}
where $F_{\mathrm{model}, i}$ is the model flux in filter $i$, $F$ is the
absolute flux (coming from the observed flux adjusted to it would be if it
were at a distance of $10\,$pc), and $\sigma$ is the combined uncertainty in
the flux and distance which can be found from the formal propagation of
uncertainties from magnitude and distance to flux with
equation \ref{eq:mag_to_flux_err} and
\ref{eq:propagation_of_err}:
\begin{equation}
    \label{eq:mag_to_flux_err}
    \sigma_{F}^{2} = \sigma_{M}^{2} \times \left[ \dfrac{\ln(10) F}{2.5} \right]^{2}
\end{equation}
and
\begin{equation}
    \label{eq:propagation_of_err}
    \sigma_{M}^{2} = \left( \dfrac{\partial M}{\partial m} \sigma_{m} \right)^2 + \left( \dfrac{\partial M}{\partial D} \sigma_{D} \right)^2
\end{equation}
where $D$ and $\sigma_D$ are the distance and the corresponding uncertainty. By
applying the distance-magnitude relation $M = m - 5\log_{10}(D)$, it simplifies to
\begin{equation}
    \label{eq:mag_err}
    \sigma_{M}^2 = \sigma^2_{m} + \left[ \dfrac{5}{\ln(10)} \right]^2 \left(\dfrac{\sigma_{D}}{D}\right)^2.
\end{equation}
Substituting in also the flux-magnitude relation, using the parametrisation
independent of the response of the photometric system: $m - m_{\mathrm{ZP}}=-2.5\log_{10}(F/F_{\mathrm{ZP}})$,
we arrive at the $\chi^2$ function to be minimized in the magnitude space weighted
by the formally propagated uncertainties:
\begin{equation}
    \label{eq:chi2}
    \chi^{2} = \left[\dfrac{2.5}{\ln(10)}\right]^2 \times \sum_{i}\left\{ \dfrac{1}{\sigma_{M, i}^2} \left[ \left(\dfrac{10}{D}\right)^{2} \times  10^{\frac{m_{i} - M_{\mathrm{model}, i}}{2.5}} - 1 \right]^2 \right\}
\end{equation}

The third method we provide is to sample the parameter space with an Markov chain
Monte Carlo method. This method is more useful in the case when the uncertainties
are large or with missing distance. The likelihood that has to be maximised takes
a similar form to Equation~\ref{eq:chi2}:
\begin{multline}
    \label{eq:likelihood}
    \mathcal{L} = -\dfrac{1}{2} \left[\dfrac{2.5}{\ln(10)}\right]^2 \times \\ \sum_{i} \left\{ 
    \dfrac{1}{\sigma_{M, i}^2} \left[ \left(\dfrac{10}{D}\right)^{2} \times 10^{\frac{m_{i} - M_{\mathrm{model}, i}}{2.5}} - 1 \right]^2
    + \ln(2\pi\sigma_i^2) \right\}.
\end{multline}

\subsection{Interstellar Reddening}
\label{sec:interstellar_reddening}
\begin{figure*}
    \centering
    \includegraphics[width=\textwidth]{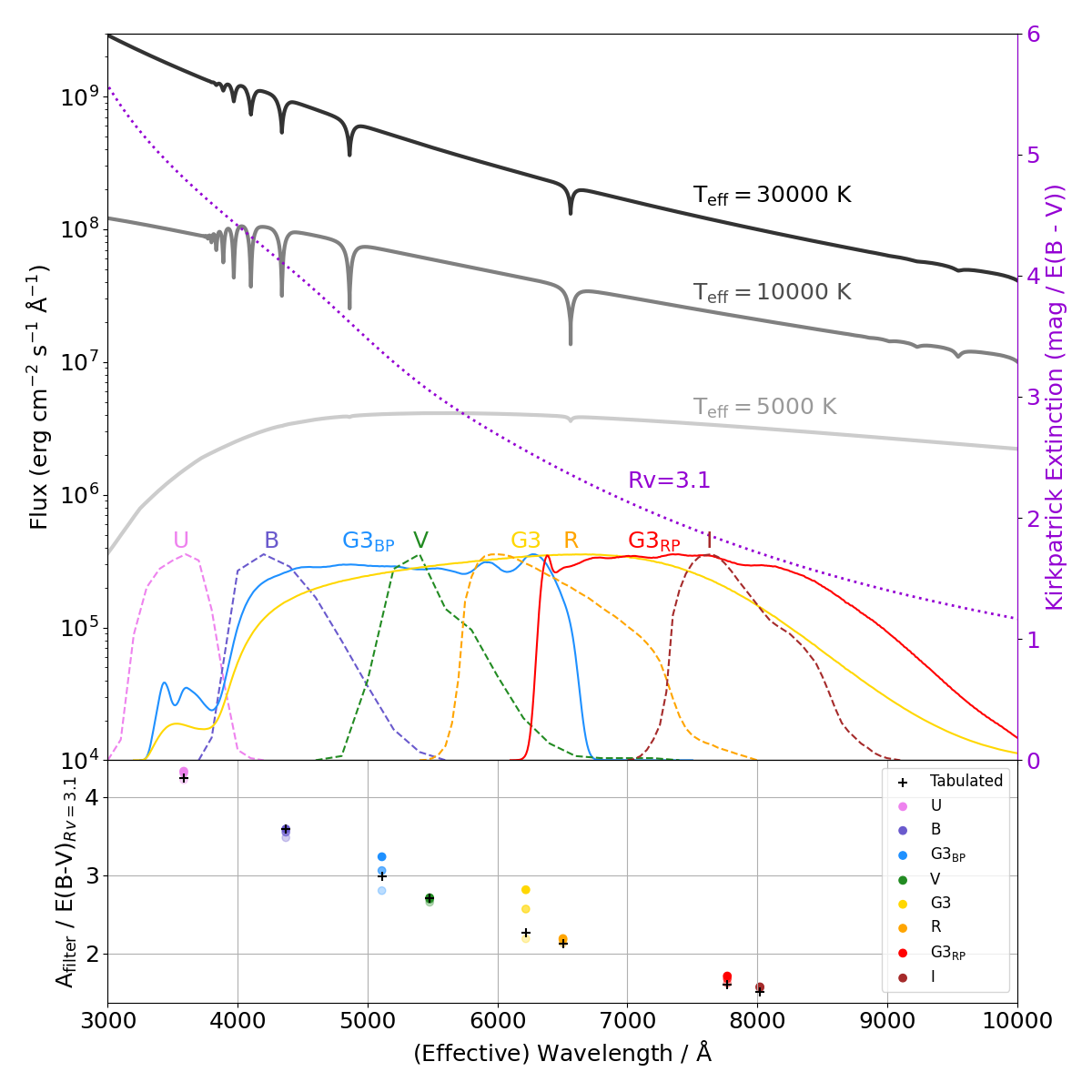}
    \caption{\textit{Top}: The grey, dark-grey and black
    lines white dwarf spectra at
    T$_{\mathrm{eff}}=5\,000, 10\,000, \mathrm{\ and\ } 30\,000$\,K. The solid
    coloured lines (with the rainbow hue sorted by the effective wavelengths)
    are the filter response function. All of filters are normalised
    independently. The purple dotted line is the Fitzpatrick extinction law.
    \textit{Bottom}: The black crosses are the pre-computed values
    from~\citet{2011ApJ...737..103S} using a synthetic spectrum of a $7\,000$\,K,
    $\log(Z) = -1.0$, and $\log(g) = 4.5$ star. The scattered points are
    extinction values obtained by convolving each of the filter profiles with
    the extinction curve and spectra. The shade of each colour follows the
    same shade as the temperature shading used in the upper plot: light shade
    is at $5\,000$\,K, medium shade is at $10\,000$\,K and the darkest shade is at
    $30\,000$\,K. It is clear that when using Gaia filters, which are very broad,
    the difference in the extinction correct is non-negligible because the
    spectral energy distribution can vary significantly within the wavelength
    range of these filters.}
    \label{fig:interstellar_reddening}
\end{figure*}

When interstellar reddening is included in the calculation, the $\chi^2$
minimisation function becomes
\begin{multline}
    \chi^{2} = \left[\dfrac{2.5}{\ln(10)}\right]^2 \times  \\
    \sum_{i}\left\{ \dfrac{1}{\sigma_{M, i}^2} \left[ \left(\dfrac{10}{D}\right)^{2} \times 10^{\frac{m_{i} - M_{\mathrm{model}, i} - A_{i}(D)}{2.5}} - 1 \right]^2 \right\}
\end{multline}

and the likelihood function to be maximised becomes
\begin{multline}
    \mathcal{L} = -\dfrac{1}{2} \left[\dfrac{2.5}{\ln(10)}\right]^2 \times \\ \sum_{i} \left\{ 
    \dfrac{1}{\sigma_{M, i}^2} \left[ \left(\dfrac{10}{D}\right)^{2} \times  10^{\frac{m_{i} - M_{\mathrm{model}, i} - A_{i}(D)}{2.5}} - 1 \right]^2
    + \ln(2\pi\sigma_i^2) \right\}.
\end{multline}

where $A_i$ is the total extinction in filter $i$ at distance $D$. In the case
when the distance is known, this is a scale value; when the distance is also
to be fitted, $A$ is not directly provided but instead, it is computed from the
$E(B-V)$ at that distance and a choice of $R_{V}$, which is defaulted to $3.1$.
The reddening vector can be applied in two ways: (1) it can be approximated by
interpolating over the effective wavelengths of the broadband
filters\footnote{CTIO $UBVRI$, UKIRT $JHKL'$, Gunn $griz$, SDSS $ugriz$,
PS1 $grizy$, LSST $ugrizy$, DES $grizY$ and WFC3 $F218W, F225W$ and $F275W$.
We note that LSST is now renamed Simonyi Survey Telescope, at the Vera Rubin
Observatory, but it was printed under the former designation in the referenced
article.} available in Table~6 of \citet{2011ApJ...737..103S}; and (2) by
treating the spectral energy distribution functions more properly, we convolve
the respective filter functions with all the DA
models~\citep{2009ApJ...696.1755T, 2010MmSAI..81..921K} available at the
Spanish Virtual
Observatory\footnote{\url{http://svo2.cab.inta-csic.es/theory/main/}} following
the instructions using equation A1 in the Appendix of \citet[][SF11]{2011ApJ...737..103S}. We have pre-computed the
reddening per unit E(B-V) from $Rv = 2.1$ to $5.1$ in $0.5$ step increment, using the
Fitzpatrick extinction law~\citep{1999PASP..111...63F} with the
\texttt{extinction}\footnote{\url{https://github.com/kbarbary/extinction}}
package~\citep{2016zndo....804967B}, for every theoretical
spectrum available over all the filters available from the Montreal synthetic
photometry table (see Figure~\ref{fig:interstellar_reddening}).
In other words, we have a third (effective temperature) and a fourth (surface gravity) dimension
of Table 6 in SF11. Our values are making use of WD spectra instead of using a single spectrum of an MS star with $\log(Z) = -1.0$, and $\log(g) = 4.5$ at $7\,000$\,K in SF11.
The synthetic
spectra do not cover some of the bluest filters, in such cases, we extend
those spectra with blackbody profiles. The grid is stored in CSV format and
is ready to be interpolated upon initialisation of this toolkit.

\subsection{Fitting Distance}
Singly evolved WDs have a surface gravity distribution that
strongly peaks at $\left<\log_{10}(g)\right> = 7.998 \pm 0.011$,
as was found empirically in the DA sample in SDSS DR16~\citep{2021MNRAS.507.4646K},
corresponding to a mean mass of $\left<\mathcal{M}_i\right> = 0.618 \pm 0.006 \msun$.
When studying a large sample of WDs,
fitting the photometric distances for the study of a population is still
useful when reporting an averaged quantity as the final results. The
distributions of the solution are, however, mostly statistical noise and not
representative to the true distribution when both the strongly degenerate
parameters: surface gravity and distance are fitted simultaneous. Equation
\ref{eq:lsq} and \ref{eq:likelihood} are reused in this case, however, the
distance modulus, which is a function of distance, becomes
a free parameter to be fitted.

\subsection{Comparison against EDR3 catalogue of WDs}
\label{sec:gf21_comparison}
The catalogue available from GF21 has presented an excellent sample of WDs for
comparison to validate our photometric fitting. There is a
slight difference between our photometric fitting methodology and that of GF21:
in this work, we are correcting for the interstellar reddening at the
instantaneous temperature and surface gravity in each iteration of the fit,
as opposed to using a global average of the extinction per
filter~($A_{G}=0.835A_{V}$, $A_{G_{\mathrm{BP}}}=0.6496A_{V}$, and
$A_{G_{\mathrm{RP}}}=1.13894A_{V}$). The underlying models
also differ slightly as the code base diverged slightly with
time\footnote{\url{https://www.astro.umontreal.ca/~bergeron/CoolingModels/}
and \url{https://warwick.ac.uk/fac/sci/physics/research/astro/people/tremblay/modelgrids}}.

As the spectral energy distribution function changes in the flux density
within the broadband filter, the total extinctions of a hotter object are
large because of more bluer photons are being extincted compared to a cooler
object. This is measureable within the wavelength range of a broadband filter.
The effect is small but not negligible\footnote{This effect
becomes even more important when the distance is a free-parameter to be fitted.
But it does not lead to degeneracy because the reddening is non-linear and
the effect is monotonic as a function of wavelength.} (see also Section
\textsection~\ref{sec:interstellar_reddening} and \textsection~\ref{sec:utility}).
It is clearly shown in Figure~\ref{fig:interstellar_reddening} that the
differences in the extinction in the Gaia filters are much larger than
those in the narrower UBVRI filters. The effect is also slightly
stronger in the blue end because of the larger change in the spectral
shape as temperature changes from $30\,000$\,K to $5\,000$\,K. Therefore, we
convolve the filter profile with the DA spectra to provide a table of
pre-computed extinction coefficient as a function of temperature and
surface gravity to allow more accurate reddening correction.

For the above reasons, some scatter is expected when the
fitted values from the two works are compared against each other.
Particularly among the hotter WDs, their colours deviate
much more from the global average than the average population
does. Their blue spectral energy distribution should be more
extincted than cooler WDs ($G_{\mathrm{BP}}$ has a much
larger spread than $G_{\mathrm{RP}})$. Otherwise, we follow as closely
to their analysis method as possible: the solutions are
fitted with the three Gaia filters, distance, and reddening with using the
catalogued values as provided from their work.
The distribution of the fitted effective temperatures are aligned well with
the identity line, the Pearson correlation coefficient of the fitted effective
temperature (for all WDs that have valid solutions fitted
regardless of the quality of fit) is $0.95$ between this and GF21. We are
only displaying fitted results above $3\,500$\,K, in line with the GF21
criteria when publishing their catalogue.

\begin{figure}
    \centering
    \includegraphics[width=\columnwidth]{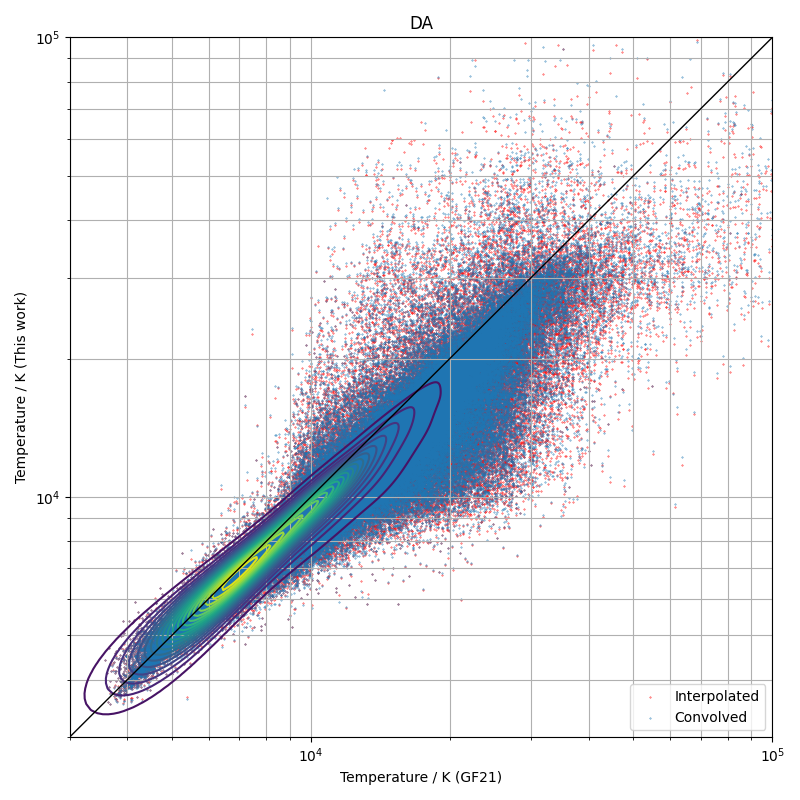}
    \caption{Comparison of the effective temperature fitting against GF21. The blue scattered points show all the fitted temperatures using our preferred reddening treatment adopting the interpolated reddening computed for each filter at each temperature available from the collection from \citet{2010MmSAI..81..921K}. The red points compares when the reddening is interpolated directly using the table from \citet{2011ApJ...737..103S} computed using a synthetic spectrum of a $7\,000$\,K  $\log(Z) = -1$, and $\log(g) = 4.5$ star. The colour lines show the density of scattered points that cannot be resolved, each level of inter-contour space contains $5\%$ of the total data. The identity line is plotted as a visual guide. The contour lines appear skewed due to the use of log-axes.}
    \label{fig:comparison}
\end{figure}

\section{White Dwarf Luminosity Function}
A WDLF is the number density of WD as a function of luminosity.
It is an evolving function with time, and it is a common tool for deriving the
age of a stellar population. Its shape and normalisation are determined
from only a few parameters. \citet{1987ApJ...315L..77W} compared an observed 
WDLF derived from the Luyten Half-Second~(LHS) catalogue with a theoretical
WDLF to obtain an estimate of the age of the Galaxy for the first time with
this technique. \citet{1990ApJ...352..605N} examined WDLFs with various SFH
scenarios. They showed that WDLF is a sensitive probe of the star formation
history~(SFH) as it shows signatures of irregularities in the SFH such as bursts
and lulls. \citet{2013MNRAS.434.1549R} took it further to address this inverse
problem mathematically and showed some success in recovering the SFH of the
solar neighbourhood when compared against SFH computed from other methods. By
decomposing the disks and halo components of the Milky Way, we can have an
independent view of the past SFH revealed by only the WD populations, where
they are most useful in deriving the SFH of old stellar
populations~\citep{2011MNRAS.417...93R, 2017ASPC..509...25L}.

he construction of a WDLF is intuitively: stars were formed in
a distribution of mass~($\mathcal{M}_i$), described by the initial mass
function~(IMF, $\phi$). Then, they spend their 
lifetime~($t_{\mathrm{MS}}$) carrying out nuclear burning, and the time they
spend depends mainly on their mass. Towards the end stage of stellar evolution
stars shed most of the atmosphere, which is modelled by the initial-final mass
relation~(IFMR, $\zeta$). Once they have become WDs, all that is left is to
know how long it has been cooling~($t_{\mathrm{cool}}$) in order to reach the
current luminosity. 

However, to encompass all these physical
processes in an integral even with pre-computed look-up table, the construction
of a theoretical WDLF is computationally expensive because of the large dynamic
range of varies quantities over the total evolution time. An integrator has to
use fine steps to achieve the required precision. The use of interpolation over
the pre-computed lookup tables has already significantly reduced the computation
time. But the repeated calling of multiple interpolate objects inside an
integral still requires a few minutes to compute a single WDLF~(at a magnitude
resolution of $0.1$\,mag) using a single typical computer processor thread.
We have carefully interpolated and integrated over the model grids, because
they are both susceptible to significant rounding errors given
the huge dynamic ranges the variables cover. However, users are reminded 
that numerical issues may still arise; for example, in the case of a
simple starburst with a duration of $10^6$\,yrs, it requires a
relative error tolerance of $10^{-10}$ in order to integrate properly for an
old population. 

It is also known that binary WD mergers lead to
an underestimation of the age of the WDs, hence they lead to contribution of
density at the ``wrong magnitude'' in a WDLF assumed to be attributed solely
by singly evolved WDs~\citep{2020A&A...636A..31T}. Another limitation is that
this work does not account for the WDs escaped from clusters and have since
drifted into the few-hundred-pc of the solar neighbourhood~(e.g.\
\citealt{2013ApJ...770..140Z,2022ApJ...926..132H, 2022ApJ...926L..24M}).

The integral for a WDLF when parameterised with bolometric
magnitude~($M_\mathrm{bol}$, as opposed to luminosity) can be written as

\begin{equation}
    n(M_{\mathrm{bol}}) = \int_{\mathcal{M}_l}^{\mathcal{M}_u}
        \tau(M_\mathrm{bol}, \mathcal{M}_f)
        \psi(T_0, M_\mathrm{bol}, \mathcal{M}_i, m, Z)
        \phi(\mathcal{M}_i) d\mathcal{M}_i
\end{equation}
where $n$ is the number density, $\tau$ is the inverse cooling rate, $\psi$ is
the relative star formation rate, $\phi$ is the initial mass function; and their
dependent variables: $M_\mathrm{bol}$ is the absolute bolometric
magnitude, $\mathcal{M}_f$ is the WD mass, $T_0$ is the look-back time, $\mathcal{M}_i$ is
the progenitor MS mass, $Z$ is the metallicity, $\mathcal{M}_l$ is the minimum
progenitor MS mass that could have singly evolved into a WD in the given time,
and $\mathcal{M}_u$ is the maximum progenitor MS mass.

The inverse cooling rate
\begin{equation}
    \tau(M_\mathrm{bol}, \mathcal{M}_f) = \dfrac{dt_{\mathrm{cool}}}{dM_\mathrm{bol}} \left( M_\mathrm{bol}, \mathcal{M}_f \right)
\end{equation}
is a quantity taken from the pre-computed grid of cooling models. 

The relative star formation rate is expressed as a function of look-back time,
\begin{align}
    &\psi(T_0, M_\mathrm{bol}, \mathcal{M}_i, \mathcal{M}_f, Z) =\\
    &\qquad\psi\left[T_0 - t_{\mathrm{cool}}\left(M_\mathrm{bol}, \mathcal{M}_f\right) - t_{\mathrm{MS}}\left(\mathcal{M}_i, Z\right)\right].
\end{align}
The absolute normalisation is not needed when the total stellar mass is coming
from observations; the theoretical WDLF only needs to multiply with a
constant (the total number density) to account for the normalisation.

The IFMR takes a simple form of
\begin{equation}
    \mathcal{M}_f = \zeta(\mathcal{M}_i),
\end{equation}
although there is evidence that more metal-rich stars lose more
envelope~\citep{2007ApJ...671..761K}, there is insufficient empirical data to
derive an IFMR at metallicity much lower or higher than solar abundance.
In the following, we will introduce the class of \texttt{WDLF}. Then, the
physical and mathematical descriptions of five different physical inputs and
their effects on a WDLF will be explained in Subsection \textsection~5.2--5.6.

\subsection{Class: \texttt{WDLF}}
\label{sec:wdlf}
This class inherits from the \texttt{AtmosphereModelReader} and the
\texttt{CoolingModelReader}.

\begin{figure*}
    \centering
    \includegraphics[width=\textwidth]{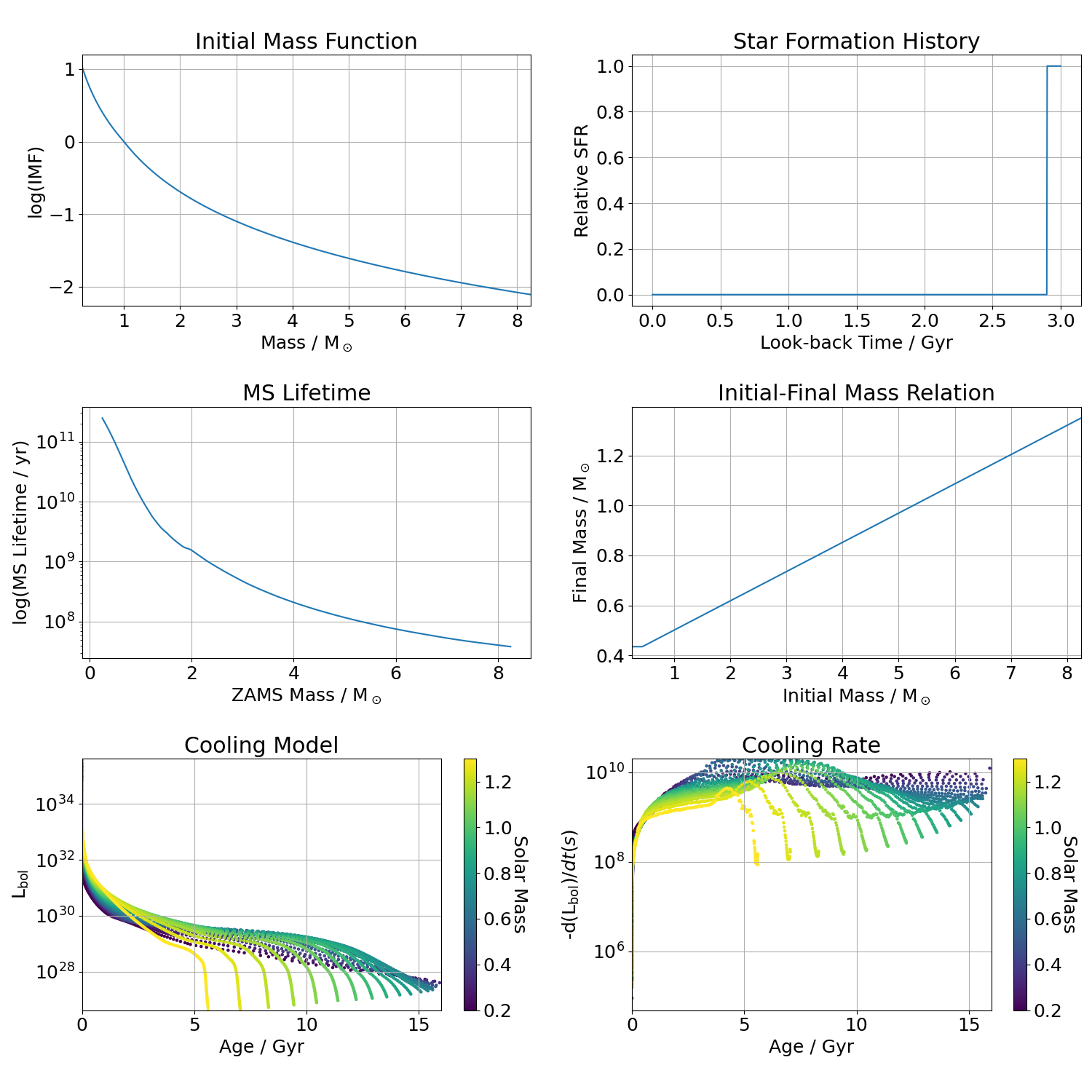}
    \caption{Top left: Input initial mass function. Top right: Input star
    formation history. Middle left: Input MS lifetime model. Middle right:
    Input IFMR model. Bottom left: Input cooling model showing the absolute
    bolometric magnitude as a function of mass and age. Bottom right: Input
    cooling model showing the rate of change of bolometric magnitude as a
    function of mass and age.}
    \label{fig:input_model}
\end{figure*}

A \texttt{WDLF} object requires seven models for computation: (1)~Star
Formation History~(SFH), (2)~Initial Mass Function~(IMF), (3)~Initial Final
Mass Relation~(IFMR), (4)~total stellar evolution time model, and (5-7) low,
intermediate and high mass WD cooling models.
A \texttt{WDLF} object comes with a method to produce
a diagnostic plot to inspect all the input models: the IMF, the SFH, the total
stellar evolution time as a function of mass, the IFMR, and the cooling model
in two formats: bolometric magnitude as a function of age, and the rate of
change of the bolometric magnitude as a function of
age~(Figure~\ref{fig:input_model} is an example of this diagnostic plot).

In the following comparisons, the default configuration uses the (1)~SFH with a
starburst of $10^8$ starting at the labelled time, (2)~IMF
from \citet{2003PASP..115..763C} for single stars, (3)~IFMR from
\citet{2008MNRAS.387.1693C}, and (4)~cooling models from the Montreal group.
While comparing the theoretical WDLFs in the following
sections. We are altering one variable at a time in the comparisons.

\textit{The WDLFs in all the figures in this section are
normalised at $10$\,mag to a log-number density of $0.0$.}

\subsection{Star Formation History}
\begin{figure}
    \centering
    \includegraphics[width=\columnwidth]{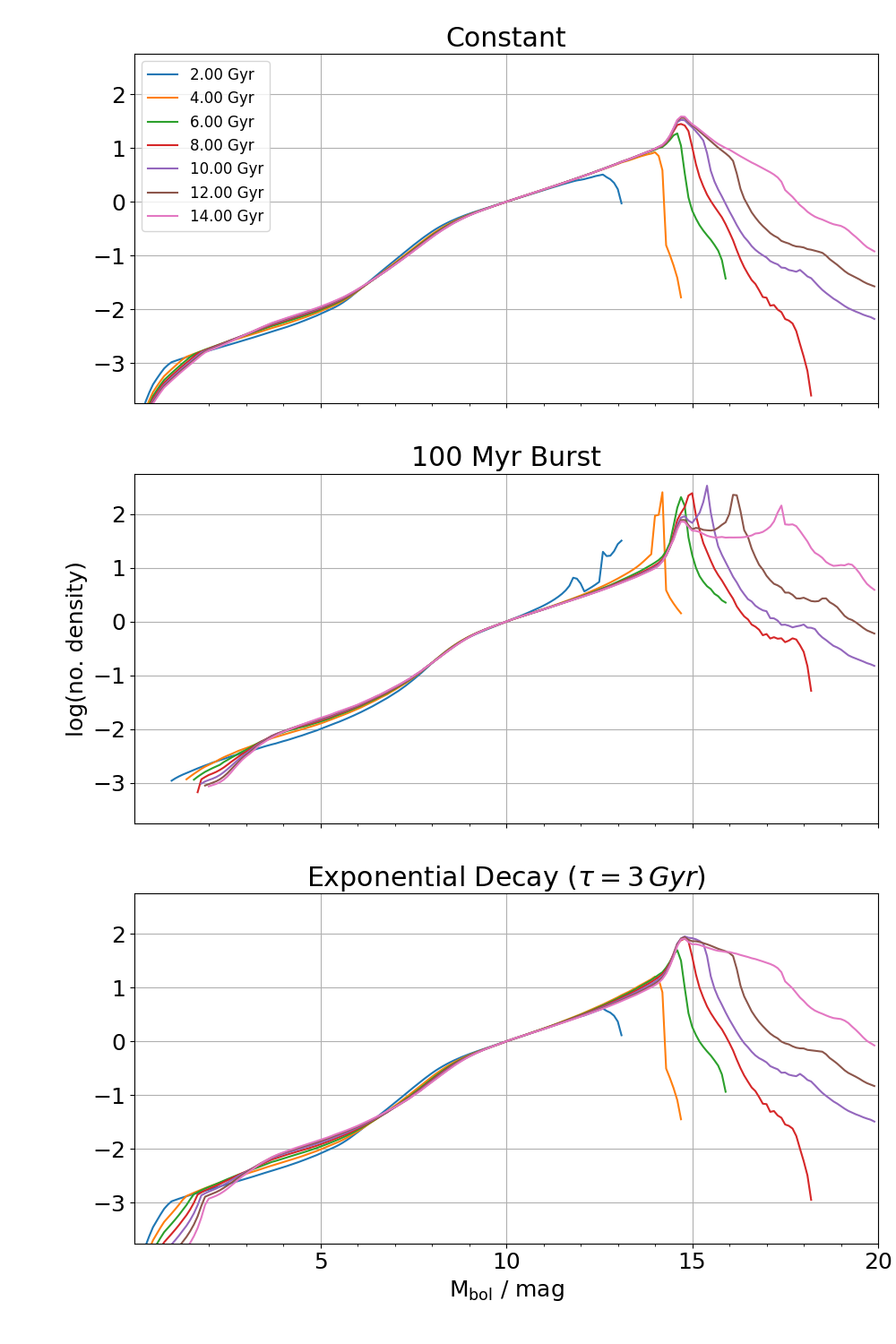}
    \caption{Comparing WDLFs with the three default types of star formation
    history from $2$ to $14\,$Gyr in increment of $2$\,Gyr. Top: constant star
    formation since the labelled lookback time. Middle: a starburst in the 
    first 100\,Myr since the labelled lookback time. Bottom: an exponentially
    decaying star formation rate with a decay constant of 3\,Gyr since the
    labelled lookback time.}
    \label{fig:compare_sfr_age_type}
\end{figure}

WDLFs with various star formation histories are shown in
Figure~\ref{fig:compare_sfr_age_type}. The bright edge of the WDLFs lines up
almost perfectly because WDs arrive at the corresponding bolometric magnitudes
at a constant rate. The gradient is most sensitive to a time-dependent
variable, for example, a change in the SFR. The sudden increase in the number
density at $M_\mathrm{bol}=14.5$ corresponds to the slow down in the cooling
due to crystallisation that releases a significant amount of heat. This bump
is the most prominent in the starburst profile at $6$\,Gyr old because WDs
with $\sim0.8-1.0\,\msun$ have the lowest cooling rates, they correspond
to $3-5\msun$ ZAMS mass that spends $0.1-0.5$\,Gyr in the MS and $5-6$\,Gyr
to reach $14.5$\,mag~($\sim 5\,500$\,K, \citealp{2019ApJ...876...67B}). The
detailed description of the shape of the WDLFs due to various cooling
mechanisms should be referred to Figure~5 and Section~3
of \citealp{2001PASP..113..409F}. These WDLFs are only served to demonstrate
the effect of SFH on their shapes, they illustrate that the faint end
of a WDLF is the most important and sensitive in revealing the past
star formation history of a stellar population.

The SFH has one of the strongest effects on the shape and normalisation of a
WDLF. We only provide three types of simple form SFH profiles. All of them are
controlled by a few simple parameters, and the fourth option is to provide a
callable function of star formation rate as a function of lookback time:
(1)~\textit{Constant} profile depends only on the age of the population,
i.e. the lookback time since the beginning of star formation,
and \textbf{not} the time since the first WD was formed. (2)~\textit{Burst}
profile which depends on the onset of star formation as well as the duration
of a constant burst of star formation. It would be useful for studying open
or globular clusters (3)~An ~\textit{exponentially decaying}
profile that is governed by the onset of star formation, the decay coefficient
and the duration of the star formation. It should give a
first good guess for a disk population. By default, it continues to decay
indefinitely. (4)~A ~\textit{manually provided callable function} that can take
any form, though users should be careful with the extrapolation setting,
a smoothly extrapolated value or zero should be returned. In
the case of \texttt{NaN} or $\pm$\texttt{inf} being returned, we set the value to zero.

This is by far the single most important time-dependent variable when
computing a WDLF. Deriving the SFH from a WDLF is one of the most important
reasons we are interested in computing them in the first place. It was proven
to successfully mathematically invert a WDLF to retrieve the SFH of the solar
neighbourhood~\citep{2013MNRAS.434.1549R} with the SDSS and SuperCOSMOS
WDLFs~\citep{2006AJ....132.1221H, 2011MNRAS.417...93R}.

Since we normalise the output WDLF by the total integrated number density at
all luminosities, the absolute normalisation of the input SFH is discarded when
provided manually.

\subsection{Initial Mass Function (IMF)}
\begin{figure}
    \centering
    \includegraphics[width=\columnwidth]{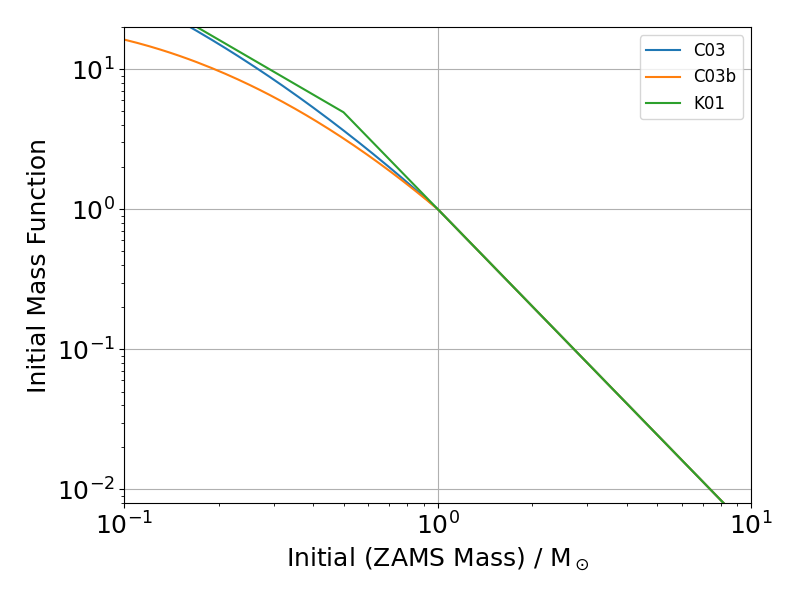}
    \caption{Comparing the three IMFs provided in this package. They are
    identical in the mass range above $1\,\msun$, while under this mass
    limit, the differences are small in the mass range ($\gtrapprox0.9\,\msun$)
    that is relevant to singly evolved MS that had enough time to evolve into a
    WD given the finite age of the Universe.}
    \label{fig:imfs}
\end{figure}

\begin{figure}
    \centering
    \includegraphics[width=\columnwidth]{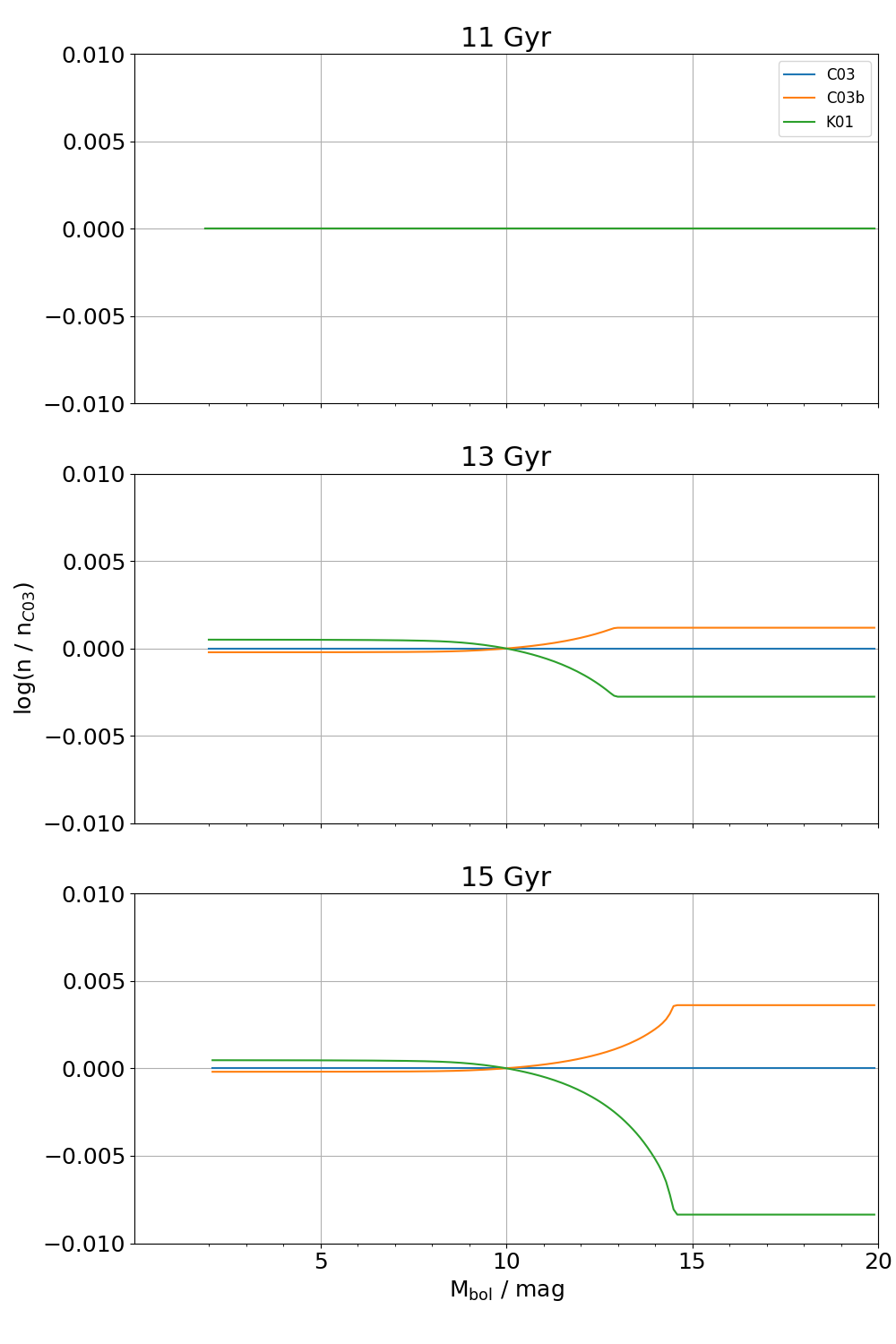}
    \caption{The initial mass functions are identical above $1\,\msun$, only the
    oldest populations show differences in the bright end of the WDLFs. Due to
    almost identical computed WDLFs, only the differences of the WDLFs
    relative to the C03 are shown.}
    \label{fig:wdlf_compare_imf}
\end{figure}

The choice of any of the three built-in Galactic IMFs has little
effect on the shape of the WDLF because they are identical above $1\,\msun$.
They only diverge slightly when the mass is below that $1\,\msun$. This effect
can appear in the bright end of a WDLF for old populations when the low mass
stars star becoming WDs~(See Figure~\ref{fig:imfs} and
\ref{fig:wdlf_compare_imf}). Apart from the three built-in IMF functions, a
callable function can be supplied as a manual IMF. For
example, if a user would like to explore the effect of a top heavy IMF, they
would have to provide it manually. This lists the four options that are
available:

\begin{enumerate}
    \item K01: \citet{2001MNRAS.322..231K} is a three-part power law that only the
    more massive two are of relevance to this work, with
    $\xi(\mathcal{M}_i) = \mathcal{M}_i^{-\alpha}$ where
    \begin{equation}
        \alpha =
        \begin{cases}
            0.3, \quad \mathcal{M}_i \leq 0.08\,\msun\\
            1.3, \quad 0.08 \leq \mathcal{M}_i \leq 0.5\,\msun\\
            2.3, \quad \mathcal{M}_i \geq 0.5\,\msun.
        \end{cases}
    \end{equation}
    \item C03: \citet{2003PASP..115..763C} shares the same power law with
    \citet{2001MNRAS.322..231K} for mass above $1\,\msun$, while below that,
    the IMF follows a log-normal form with
    \begin{equation}
        \xi(\mathcal{M}_i) = \dfrac{0.158^{+0.051}_{-0.046}}{\mathcal{M}_i \ln(10)} \times
            \exp{\left\{\dfrac{\left[\log_{10}(\mathcal{M}_i) - \log_{10}\left(0.079^{-0.016}_{+0.021}\right)\right]^2}{2 \times (0.69^{-0.01}_{+0.05})^2}\right\}}
    \end{equation}
    \item C03b: \citet[][including binary]{2003PASP..115..763C} also consider the correction for binaries which gives a much shallower IMF at the low mass end
    \begin{equation}
        \xi(\mathcal{M}_i) = \dfrac{0.086}{\mathcal{M}_i \ln(10)} \times
            \exp{\left\{\dfrac{\left[\log_{10}(\mathcal{M}_i) - \log_{10}\left(0.022\right)\right]^2}{2 \times (0.57)^2}\right\}}
    \end{equation}
    \item Manually provide an interpolated function that returns the initial mass density as a list or an array (even if a single point is returned).
\end{enumerate}

There is a very weak dependency on the IMF because the three IMFs are only
different at mass below $1\,\msun$, which has an MS lifetime of
$\sim11$\,Gyr for solar metallicity stars. Hence, we only
start to see a difference in the WDLF at ages above
$11$\,Gyr~(Figure~\ref{fig:wdlf_compare_imf}). The IMFs only diverge
significantly well into the M-dwarf/brown dwarf
regime. Even at a computed age of $15$\,Gyr
we are only including stars with ZAM masses as low as $\sim$ $0.96\,\msun$.
Nevertheless, if a metal-poor MS lifetime model is used, this can change the
picture slightly as they are more short-lived.

\subsection{Initial-Final Mass Relations}
\begin{figure*}
    \centering
    \includegraphics[width=\textwidth]{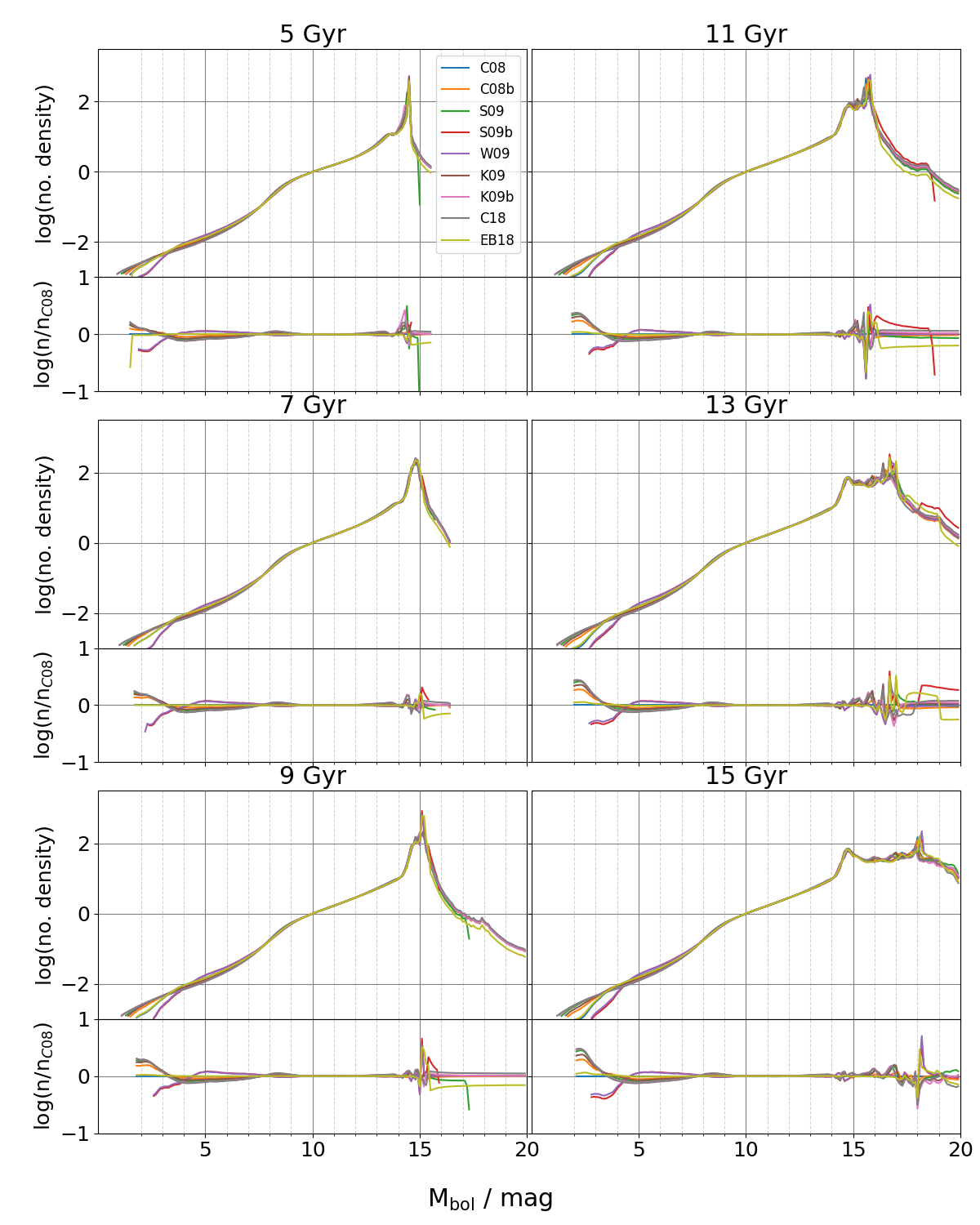}
    \caption{Comparing WDLFs with different input IFMRs. Its effect on the WDLF
    is small because of the similarity between them. The large difference at
    the bright end is due to the relatively large difference in the cooling
    rate among the new hot WDs. The differences in the faint end come from the
    accumulation of different cooling rates at slightly different WD masses for
    a progenitor at a given mass.}
    \label{fig:wdlf_compare_ifmr}
\end{figure*}

In Fig.~\ref{fig:wdlf_compare_ifmr}, the WDLFs in the range of $9$--$14$\,mag
line up almost perfectly at any age. This is because WDs with a `typical mass'
do not have strong mass-dependent physical phenomenon that affects the WD
evolution. They all cool at roughly the same rate across the mass range of the
models. However, the difference is still measurable and obvious by eye when
the systems have sufficient time to evolve and for the WDs to cool. In 
Figure~\ref{fig:wdlf_compare_ifmr}, the $5$\,Gyr comparison plot is exhibiting
the dip in the EB18 IFMR just above $5\,\msun$ at $14.5$\,mag. Meanwhile the strong
feature at $14.0$--$14.5$\,mag in S09~(red), K09~(brown), K09b~(pink), and
C18~(grey) are when the respective IFMRs are crossing the C09 IFMR from a higher
final mass than C08's to a lower final mass. Such an effect can be seen propagating
down the WDLFs from the younger populations to the older at about $1$\,mag every
$2$\,Gyr. Similarly, in the oldest system, we can see that when $0.9\,\msun$
have recently become WDs in the $13$ and $15$\,Gyr, only the S09b~(red) and
W09~(purple) models that return lower WD mass than that from C08 are showing
deficiencies at the brightest end. 

The comparison in Figure~\ref{fig:wdlf_compare_ifmr} reaffirms
the finding from \citet{2008MNRAS.387.1693C} that
the choice of IFMR has little effect on the
shape of the WDLFs at $\mathrm{M}_{\mathrm{bol}}>4$.
We have also shown in the figure that the peaks of the WDLFs differ slightly
in the bolometric magnitude at a given age. That comes from the
accumulation of a slightly different cooling rate over a cosmic time when a
progenitor at a given mass turns into slightly different mass WDs as prescribed
by the different IFMRs.
 
Human civilisation has existed for a mere few thousand years, and modern
astronomy (with digital aid) for no more than a hundred. We
do not have direct observations on the total mass loss of a star at the end
stage of stellar evolution. It depends on observations of WDs and giants in
clusters and iteratively comparing with stellar evolution models. There are a
number of IFMRs available from studying globular
clusters~\citep{2004A&A...420..515M, 2009ApJ...705..408K}, open
clusters~\citep{2009ApJ...693..355W, 2016ApJ...818...84C}, a mix of globular
and open cluster~\citep{2018ApJ...866...21C}, wide MS-WD
binaries~\citep{2008A&A...477..213C, 2012ApJ...746..144Z}, wide
turnoff/subgiant-WD binaries~\citep{2021ApJ...923..181B}, wide WD-WD
binaries~\citep{2015ASPC..493..325C, 2015ApJ...815...63A}, and colour-magnitude
diagram fitting~\citep{2018ApJ...860L..17E}. We have chosen six (plus three
with two-part fitting) works that have a good mass coverage to be included
in this work, and just like the other functions above, a manually provided
callable function is also accepted:

\begin{enumerate}
    \item C08: \citet{2008MNRAS.387.1693C} -- this work reanalysed all the known WDs in clusters and MS-WD binaries homogeneously by using a single stellar evolution model and a single WD cooling model. The sample covers a mass range of $1.5-6.4\,\msun$, giving a best-fit IFMR of
    \begin{equation}
        \mathcal{M}_f = (0.117 \pm 0.004)\,\mathcal{M}_i + (0.384 \pm 0.011)\,\msun.
    \end{equation}
    \item C08b: \citet[][two-part]{2008MNRAS.387.1693C} -- in the same work, they also fitted with a 2-part IFMR with a breakpoint at $2.7\,\msun$, which gives a steeper relation at the high mass end with
    \begin{equation}
        \mathcal{M}_f = \begin{cases}
                  (0.096 \pm 0.005)\,\mathcal{M}_i + (0.429 \pm 0.015)\,\msun,\\
                  \qquad(\mathcal{M}_i \leq 2.7\,\msun)\\
                  (0.137 \pm 0.007)\,\mathcal{M}_i + (0.318 \pm 0.018)\,\msun,\\
                  \qquad(\mathcal{M}_i \geq 2.7\,\msun)
              \end{cases}
    \end{equation}
    \item S09: \citet{2009ApJ...692.1013S} -- by using 10 (young) open clusters, this relation is probing the high mass end of the population that can turn into WDs. They, however, do not have enough time for the lower mass stars to evolve into WDs hence this relation is unconstrained below $1.7\,\msun$ and we extrapolate if the given initial mass is below $1.7\,\msun$. The upper limit of the data set is $8.5\,\msun$:
    \begin{equation}
        \mathcal{M}_f = 0.084 \mathcal{M}_i \pm 0.466\,\msun.
    \end{equation}
    \item S09b: \citet[][two-part]{2009ApJ...692.1013S} -- in the same work, they also fitted with a 2-part IFMR with a break point at $4\,\msun$.
    \begin{equation}
        \mathcal{M}_f = \begin{cases}
                  0.134\,\mathcal{M}_i + 0.331\,\msun, &1.7\,\msun \leq \mathcal{M}_i \leq 4.0\,\msun\\
                  0.047\,\mathcal{M}_i + 0.679\,\msun, &\mathcal{M}_i \geq 4.0\,\msun
              \end{cases}
    \end{equation}
    \item W09: \citet{2009ApJ...693..355W} is an extension to \citet{2009ApJ...692.1013S} by including M41 to the collection of open clusters, covering a reliable mass range of $1.25-8.0\,\msun$,
    \begin{equation}
        \mathcal{M}_f = (0.129 \pm 0.004) \mathcal{M}_i + (0.339 \pm 0.015)\,\msun.
    \end{equation}
    \item K09: \citet{2009ApJ...705..408K} reanalysed with all the available open clusters covering a mass range of $1.1-6.5\,\msun$ to give a relation of
    \begin{equation}
        \mathcal{M}_f = (0.109 \pm 0.007) \mathcal{M}_i + (0.428 \pm 0.025)\,\msun.
    \end{equation}
    \item K09b: \citet[][extended]{2009ApJ...705..408K} includes also the WDs from the globular cluster, M4, to give a shallower relation.
    \begin{equation}
        \mathcal{M}_f = (0.101 \pm 0.006) \mathcal{M}_i + (0.463 \pm 0.018)\,\msun.
    \end{equation}
    However, having only metal richer open cluster data at the high mass end and only the metal-poor globular cluster data at the low mass end is likely to lead to a shallower fit expected due to the metallicity dependency in the total mass loss at the late stage of stellar evolution.
    \item C18: \citet{2018ApJ...866...21C} has fitted and compared various theoretical and empirical models, we are only providing the 3-part-solution that the authors preferred, which is based on the MIST stellar evolution model,
    \begin{equation}
        \mathcal{M}_f = \begin{cases}
                  (0.080 \pm 0.016)\,\mathcal{M}_i + (0.489 \pm 0.030)\,\msun,\\
                  \qquad(0.83\,\msun \leq \mathcal{M}_i \leq 2.85\,\msun)\\
                  (0.187 \pm 0.061)\,\mathcal{M}_i + (0.184 \pm 0.199)\,\msun,\\
                  \qquad(2.85\,\msun \leq \mathcal{M}_i \leq 3.6\,\msun)\\
                  (0.107 \pm 0.016)\,\mathcal{M}_i + (0.471 \pm 0.077)\,\msun,\\
                  \qquad(3.6\,\msun \leq \mathcal{M}_i \leq 7.2\,\msun)\\
              \end{cases}
    \end{equation}
    \item EB18: \citet{2018ApJ...860L..17E} reports the relations by fitting the colour-magnitude diagram of the clean Gaia DR2 WD sequence. It is a 5-point piecewise-linear function going through the following (initial mass, final mass) coordinates: $(0.95, 0.5^{+0.01}_{-0.01})$, $(2.75^{+0.36}_{-0.31}$, $0.67^{+0.02}_{-0.02})$, $(3.54^{+0.55}_{-0.43}$, $0.81^{+0.03}_{-0.03})$, $(5.21^{+1.06}_{-0.71}$, $0.91^{+0.10}_{-0.03})$ and $(8.0, 1.37^{+0.06}_{-0.21})$.
    \item Manually provided an interpolated function that returns the final mass as a list or an array (even if a single point is returned).
\end{enumerate}

\begin{figure}
    \centering
    \includegraphics[width=\columnwidth]{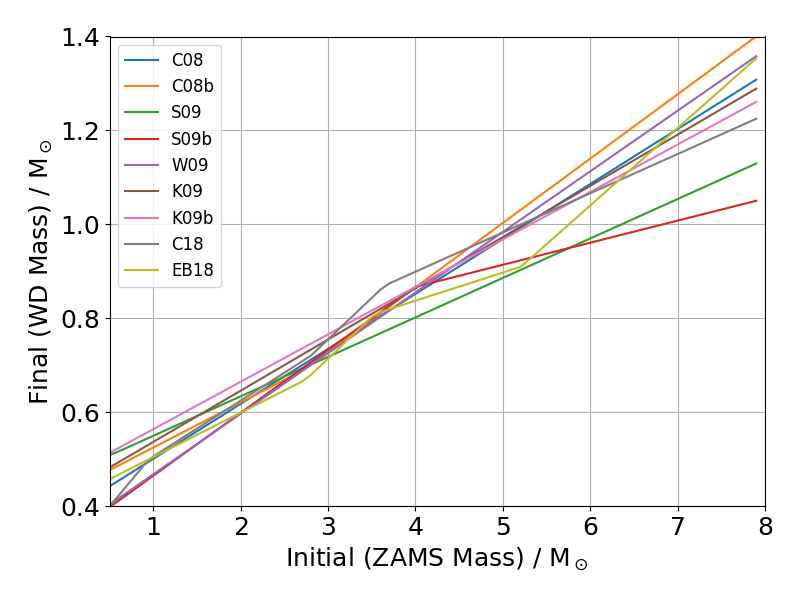}
    \caption{Comparing the three IFMRs provided in this package. They follow
    a similar trend and they are mostly lying on top of each other. S09 and S09b
    have the shallowest IFMRs at the massive end of the initial mass among all.
    However, given the steep decline of the initial mass function, $7\,\msun$
    MS stars are about 10 times less common than $3\,\msun$ MS stars and about
    100 times less common than $1\,\msun$ MS stars. Hence, the discrepancy also
    only attribute to a small effect in the final WDLFs~(See
    Figure~\ref{fig:wdlf_compare_ifmr})}.
    \label{fig:ifmrs}
\end{figure}

\subsection{Total Stellar Evolution Time}
\label{sec:evolution_time}
The effect of metallicity on the WDLF is qualitatively
described as follows: the more massive stars become WDs at roughly the same
time, so the faint end of the WDLFs show little differences. However, the
evolution times for the lower mass stars reduce with metallicity. Thus, at the
bright end of a WDLF which is dominated by low mass stars, using a low
metallicity stellar evolution model would `squeeze' the WDLF towards its
fainter part, giving a WDLF with a steeper gradient on the bright side.
This makes a system appears older if a low metallicity model is used for
interpreting a more metal-rich population, and vice versa. This effect should
be considered in conjunction with the metallicity dependence of the IFMR in
the future when such a relation becomes better constrained.

The total stellar evolution lifetime has a strong effect on the bright end of a
WDLF because the hot WDs spent most of their time since star formation as their
progenitors. However, the MS lifetime has decreasing impact on the WDLF as we
move towards the fainter end where WDs cooling time dominates over the MS
lifetime. There is also a metallicity dependency on the MS lifetime. From the
PARSEC models~\citep{2013EPJWC..4303001B}, an extremely metal poor $1\,\msun$ star
with  $Z=0.0001$ and a solar-metallicity star with $Z=0.017$ take $\sim$\,$6$\,Gyr
and $\sim$\,$11$\,Gyr to go through the stellar evolution stages, respectively,
before becoming a WD. In the \textsc{WDPhotTools}, we provide $15$ models from
PARSEC~\citep{2012MNRAS.427..127B, 2013EPJWC..4303001B, 2014MNRAS.444.2525C},
$3$ from the Geneva Code~\citep{2012A&A...537A.146E, 2012A&A...541A..41M,
2013A&A...553A..24G, 2013A&A...558A.103G}, and $15$ from the
MIST~\citep{2011ApJS..192....3P, 2013ApJS..208....4P, 2015ApJS..220...15P,
2016ApJS..222....8D, 2016ApJ...823..102C}. The following stellar evolution
lifetime models are interpolated and the option of providing a callable
function is also a possible form as input:

\begin{enumerate}
    \item PARSECz00001 (Z = $0.0001$, Y $= 0.249$)
    \item PARSECz00002 (Z = $0.0002$, Y $= 0.249$)
    \item PARSECz00005 (Z = $0.0005$, Y $= 0.249$)
    \item PARSECz0001 (Z $= 0.001$, Y $= 0.25$)
    \item PARSECz0002 (Z $= 0.002$, Y $= 0.252$)
    \item PARSECz0004 (Z $= 0.004$, Y $= 0.256$)
    \item PARSECz0006 (Z $= 0.006$, Y $= 0.259$)
    \item PARSECz0008 (Z $= 0.008$, Y $= 0.263$)
    \item PARSECz001 (Z $= 0.01$, Y $= 0.267$)
    \item PARSECz0014 (Z $= 0.014$, Y $= 0.273$)
    \item PARSECz0017 (Z $= 0.017$, Y $= 0.279$)
    \item PARSECz002 (Z $= 0.02$, Y $= 0.284$)
    \item PARSECz003 (Z $= 0.03$, Y $= 0.302$)
    \item PARSECz004 (Z $= 0.04$, Y $= 0.321$)
    \item PARSECz006 (Z $= 0.06$, Y $= 0.356$)
    \item GENEVAz002 (Z $= 0.002$)
    \item GENEVAz006 (Z $= 0.006$)
    \item GENEVAz014 (Z $= 0.014$)
    \item MISTFem400 ([Fe/H] $= -4.0$)
    \item MISTFem350 ([Fe/H] $= -3.5$)
    \item MISTFem300 ([Fe/H] $= -3.0$)
    \item MISTFem250 ([Fe/H] $= -2.5$)
    \item MISTFem200 ([Fe/H] $= -2.0$)
    \item MISTFem175 ([Fe/H] $= -1.75$)
    \item MISTFem150 ([Fe/H] $= -1.5$)
    \item MISTFem125 ([Fe/H] $= -1.25$)
    \item MISTFem100 ([Fe/H] $= -1.0$)
    \item MISTFem075 ([Fe/H] $= -0.75$)
    \item MISTFem050 ([Fe/H] $= -0.5$)
    \item MISTFem025 ([Fe/H] $= -0.25$)
    \item MISTFe000 ([Fe/H] $= 0.0$)
    \item MISTFe025 ([Fe/H] $= 0.25$)
    \item MISTFe050 ([Fe/H] $= 0.5$)
    \item Manual
\end{enumerate}

\begin{figure}
    \centering
    \includegraphics[width=\columnwidth]{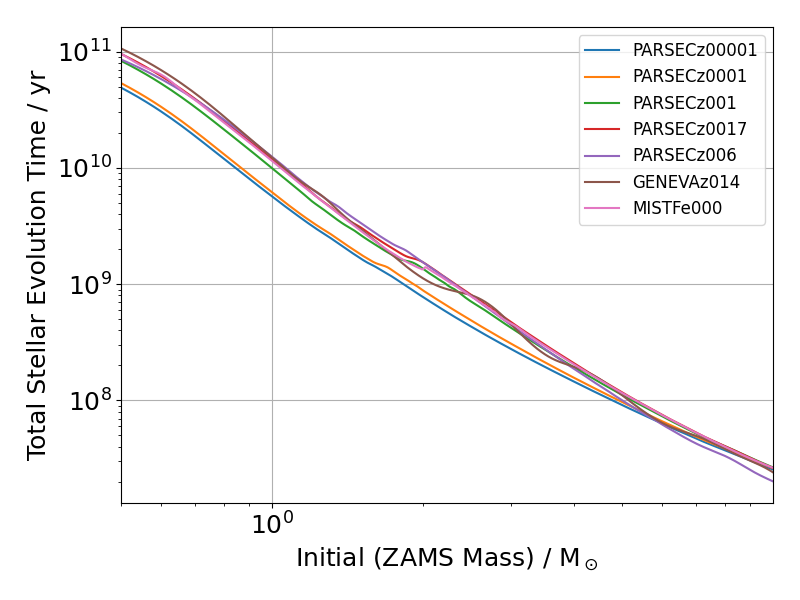}
    \caption{Comparing seven of the 33 available models: PARSEC is shown with
    metallicity at $z=0.0001, 0.001, 0.01, 0.017$ (i.e. solar) and $0.06$; Geneva
    and MIST solar metallicity model is plotted for comparison. The low
    metallicity stars evolve more quickly than their higher metallicity counterparts.
    At 1 solar mass, the most metal-poor model evolves twice as fast as the
    solar metallicity one. At higher mass, the discrepancy is smaller at
    different metallicity and from different models.}
    \label{fig:total_stellar_lifetime}
\end{figure}

A few models are plotted in Fig.~\ref{fig:total_stellar_lifetime} to
illustrate the differences in the MS lifetime at difference masses and
metallicities.

\subsection{Cooling Models}
\label{sec:cooling_models}
Most of the internal energy of a WD is the residual heat from the progenitor
once it passed the planetary nebula phase. However, there are various physical
processes that can provide an appreciable amount of energy and govern the
cooling rate of a WD at different stages. Following the time sequence in which
the physical processes that have direct effects on the photo-luminosity:
\begin{enumerate}
    \item in the first $10^8-10^9$ years, \textit{shell burning of hydrogen} via pp-chain can contribute up to $30\%$ of the total luminosity~\citep{2010ApJ...717..183R}.
    \item \textit{Neutrino losses} -- contribute to a significant fraction of energy loss in the early time of WDs when they were still hot, in the case of massive WDs, neutrino bremsstrahlung effect must also be taken into account \citep{1994ApJ...425..222H, 1996ApJS..102..411I}.
    \item \textit{Gravitational
settling} of\ $^{22}$Ne in intermediate to massive WDs releases sufficient
gravitational potential energy to prolong the cooling
times~\citep{2002ApJ...580.1077D, 2008ApJ...677..473G, 2010ApJ...719..612A}.
The heavier\ $^{22}$Ne relative to the environment that is dominated by carbon,
oxygen and nitrogen leads to a slow settling towards the core. This effect is
the most obvious in the old and metal-rich systems, such as NGC
6791~\citep{2010Natur.465..194G, 2008ApJ...678.1279B}.
    \item In the late time of
the WD evolution, convection plays a significant role in slowing down the
cooling. As temperature decreases, the convective zone grows deeper into the
interior and eventually reaches the degenerate core~(see Figure~11
from~\citealt{2010A&ARv..18..471A}). This efficiently replenishes the energy
radiated away from the photosphere, thus this process known as the
\textit{convective coupling}, modifies the relations between the WD luminosity
and core temperature~\citep{1989ApJ...347..934D, 2001PASP..113..409F}.
    \item \textit{Crystallisation} occurs as the non-degenerate ions evolve from gas
to fluid and eventually solid. The liquid-solid transition releases latent
heat that slows down the cooling process. This also couples with the release
of gravitational energy associated with changes in the carbon-oxygen
profile~\citep{1997ApJ...486..413S} when the heavier oxygen-rich crystals
displace carbon as a result of gravitational settling.
    \item Depending on the
changes in the carbon-oxygen abundance profile, and the choice of phase
diagram of a carbon-oxygen mixture, it modifies the rate of cooling and this
specific effect is colloquially known as the \textit{Phase Separation}
effect.
    \item \textit{Coulomb Interactions} modify the thermodynamical
properties of the ionic gas, in particular the specific heat. Its strength is
determined by the Coulomb coupling parameters. At first, the parameter is
small, it slowly increases as a WD cools and the ions begin to change from
gas to liquid and eventually form lattice. This releases latent heat that
contributes to $\sim$\,$5\%$ of the total luminosity~\citep{1976A&A....51..383S}.
At late time, a few modes of the lattice are excited, the heat capacity drops
according to the Debye law, which results in enhanced cooling. This process
kicks in after $10^9$\,yr for a $1.0\,\msun$ WD, and after more than a Hubble time for a
$0.5\,\msun$ WD.
\end{enumerate}

We have included $24$ cooling models from sixteen pieces of work
that cover different parts of the parameter space. The following lists the
keywords for choosing the models:

\begin{enumerate}
    \item montreal\_co\_da\_20~\citep{2020ApJ...901...93B}
    \item montreal\_co\_db\_20~\citep{2020ApJ...901...93B}
    \item lpcode\_he\_da\_07~\citep{2007MNRAS.382..779P}
    \item lpcode\_he\_da\_09~\citep{2009ApJ...704.1605A}
    \item lpcode\_co\_da\_07~\citep{2007MNRAS.382..779P}
    \item lpcode\_co\_da\_10\_z001~\citep{2010ApJ...717..183R}
    \item lpcode\_co\_da\_10\_z0001~\citep{2010ApJ...717..183R}
    \item lpcode\_co\_da\_15\_z00003~\citep{2015A&A...576A...9A}
    \item lpcode\_co\_da\_15\_z0001~\citep[]{2015A&A...576A...9A}
    \item lpcode\_co\_da\_15\_z0005~\citep[]{2015A&A...576A...9A}
    \item lpcode\_co\_db\_17\_z00005~\citep[]{2017A&A...597A..67A}
    \item lpcode\_co\_db\_17\_z0001~\citep[]{2017A&A...597A..67A}
    \item lpcode\_co\_db\_17~\citep{2017ApJ...839...11C}
    \item lpcode\_one\_da\_07~\citep{2007A&A...465..249A}
    \item lpcode\_one\_da\_19~\citep{2019A&A...625A..87C}
    \item lpcode\_one\_db\_19~\citep{2019A&A...625A..87C}
    \item lpcode\_da\_22~\citep{2013A&A...557A..19A, 2016ApJ...823..158C, 2019A&A...625A..87C}
    \item lpcode\_db\_22~\citep{2017ApJ...839...11C, 2019A&A...625A..87C}
    \item basti\_co\_da\_10~\citep{2010ApJ...716.1241S}
    \item basti\_co\_db\_10~\citep{2010ApJ...716.1241S}
    \item basti\_co\_da\_10\_nps~\citep{2010ApJ...716.1241S}
    \item basti\_co\_db\_10\_nps~\citep{2010ApJ...716.1241S}
    \item mesa\_one\_da\_18~\citep{2018MNRAS.480.1547L}
    \item mesa\_one\_db\_18~\citep{2018MNRAS.480.1547L}
\end{enumerate}

See Table~\ref{tab:cooling_models} for the details of each model. We divide
the models into three groups:
low~($\mathcal{M}/\msun < 0.5$),
intermediate~($0.5 < \mathcal{M}/\msun < 1.0$) and
high mass~($\mathcal{M}/\msun > 1.0$). For the lowest mass one, it is only
used in the cases where the IFMR gives WDs reaching such mass, and a
population that can be old enough to produce these low mass singly evolved
WDs.

\begin{table*}
    \centering
    \begin{tabular}{cccccccc}
        Reference             &    Low     & Intermediate &    High    &  Core & Atmosphere &           Mass Range $\left(\mathcal{M}_f/\msun\right)$ & Extra Notes \\\hline\hline

        \multicolumn{8}{c}{Montreal smooth grid} \\\hline
        \citet[][B20]{2020ApJ...901...93B} & \checkmark &  \checkmark  & \checkmark &    CO &     \textcolor{black}{H/He} &            $0.2-1.3$             & -- \\
        &&&&&&&\\

        \multicolumn{8}{c}{BaSTI smooth grid} \\\hline
        \citet[][S10]{2010ApJ...716.1241S}&     --     &  \checkmark  & \checkmark &    CO &      \textcolor{black}{H/He} &           $0.54-1.2$             & W/Wo$^{\dagger}$ Phase Separation\\
        &&&&&&&\\

        \multicolumn{8}{c}{LPCODE smooth grid} \\\hline
        {\citet{2013A&A...557A..19A}} & \checkmark &     --     &     --     & He & \textcolor{black}{H} & -- & -- \\
        \citet{2016ApJ...823..158C}   &     --     & \checkmark &     --     & He & \textcolor{black}{H} & -- & -- \\
        {\citet{2019A&A...625A..87C}} &     --     &     --     & \checkmark & He & \textcolor{black}{H} & -- & -- \\
        \citet{2017ApJ...839...11C}   &     --     & \checkmark &     --     & He & \textcolor{black}{He} & -- & -- \\
        {\citet{2019A&A...625A..87C}} &     --     &     --     & \checkmark & He & \textcolor{black}{He} & -- & -- \\
        &&&&&&&\\

        \multicolumn{8}{c}{LPCODE} \\\hline
        \citet{2007MNRAS.382..779P} & \checkmark &      --      &     --     & He/CO &         \textcolor{black}{H} &          $0.187-0.448$           & -- \\
        \citet[][A09]{2009ApJ...704.1605A} & \checkmark &      --      &     --     &    He &         \textcolor{black}{H} &          $0.220-0.521$           & -- \\
        \citet[][R10]{2010ApJ...717..183R} &     --     &  \checkmark  &     --     &    CO &         \textcolor{black}{H} &          $0.505-0.934$           & $Z=0.001-0.01$ \\
        {\citet{2015A&A...576A...9A}} &     --     &  \checkmark  &     --     &    CO &         \textcolor{black}{H} &          $0.506-0.826$           & $Z=0.0003-0.001$ \\
        {\citet{2017A&A...597A..67A}} & \checkmark &  \checkmark  &     --     &    CO &         \textcolor{black}{H} &          $0.434-0.838$           & $Y=0.4$ \\
        \citet{2017ApJ...839...11C} &     --     &  \checkmark  &     --     &    CO &         \textcolor{black}{He} &           $0.51-1.0$             & -- \\
        {\citet{2007A&A...465..249A}} &     --     &      --      & \checkmark &   ONe &         \textcolor{black}{He} &           $1.06-1.28$            & -- \\
        {\citet[][C19]{2019A&A...625A..87C}}&     --     &      --      & \checkmark &   ONe &      \textcolor{black}{H/He} &           $1.10-1.29$            & -- \\
        &&&&&&&\\

        \multicolumn{8}{c}{MESA-based Models} \\\hline
        \citet{2018MNRAS.480.1547L} &     --     &      --      & \checkmark & CO/Ne &      \textcolor{black}{H/He} &          $1.012-1.308$           & --

    \end{tabular}
    \caption{The complete listing of all the cooling models
    available from the public domain contained in \textsc{WDPhotTools}. They
    broadly come from four sources, where the LPCODE has the most varied
    grid for various parameters. The three \textit{smooth grid}s can provide
    the smoothest interpolated cooling models, the mix-and-match of other
    models are likely to produce artefacts. We recommend separating the
    low, intermediate and high mass models to generate subset of the WDLFs
    and add their contribution together to achieve better solutions.} The
    checkmarks in the low ($\mathcal{M}_f/\msun < 0.5$), intermediate
    ($0.5 < \mathcal{M}_f/\msun < 1.0$) and high mass ($1.0 < \mathcal{M}_f/\msun$)
    ranges denotes if the models are used for computation in that range.
    ($^{\dagger}$\textit{With/Without.})
    \label{tab:cooling_models}
\end{table*}

\begin{figure*}
    \centering
    \includegraphics[width=\textwidth]{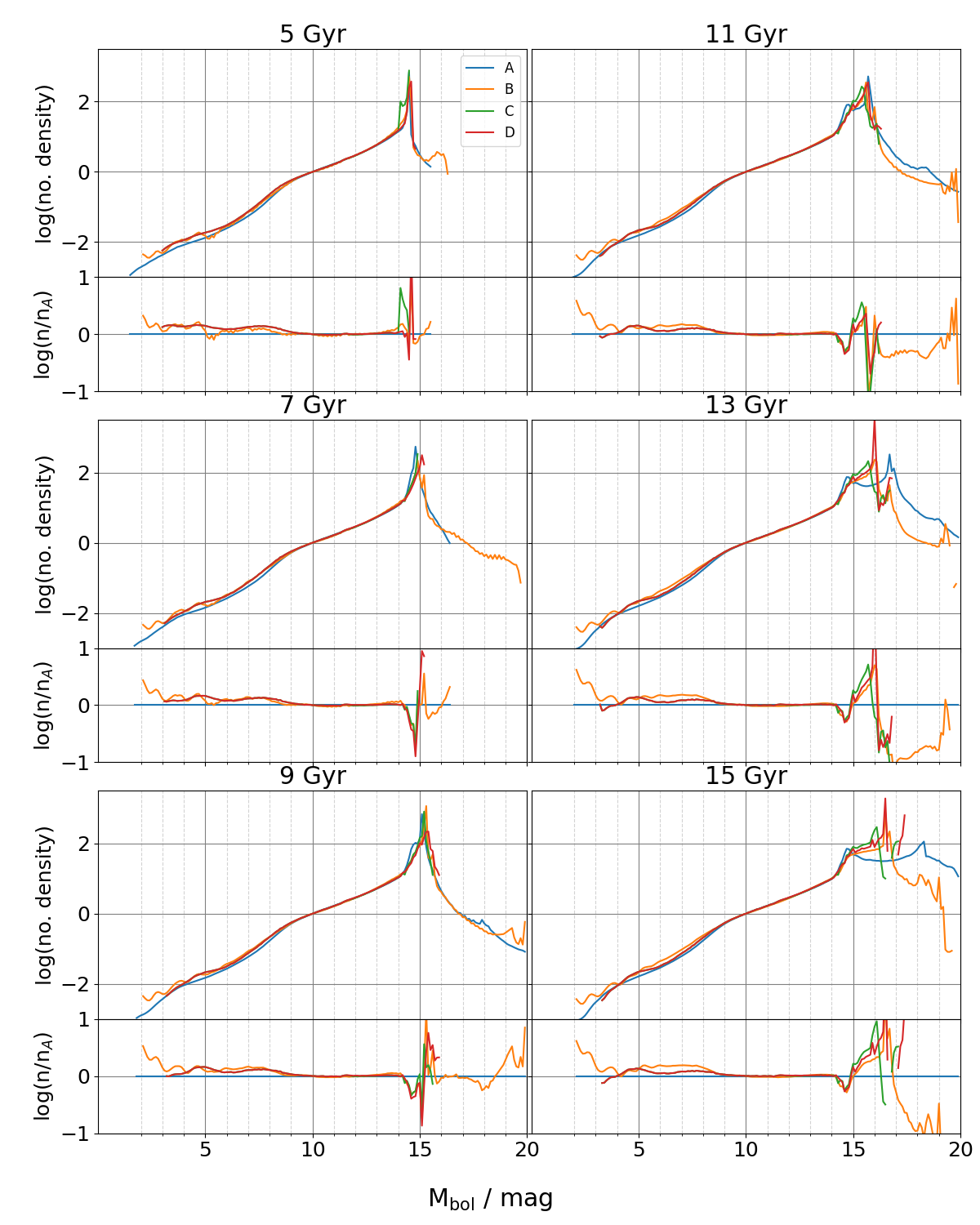}
    \caption{Comparing WDLFs with different input cooling models. See
    Table~\ref{tab:cooling_model_combination} for the choice of input models.
    The differences in the WDLFs are small down to $14$\,mag. After that, the
    input Physics in handling gravitational settling of heavy elements,
    convective coupling, crystallisation, phase separation, and Coulomb
    interactions vary and are setting in at different times (see
    Section~\ref{sec:cooling_models} for more details).
    }
    \label{fig:wdlf_compare_da_cooling_models}
\end{figure*}

There are too many possible combinations of cooling models
to display them all, so in
Figure~\ref{fig:wdlf_compare_da_cooling_models}, only four example
sets of models are shown. The combinations are listed in
Table~\ref{tab:cooling_model_combination}. In order to assess the quality of
the interpolation at the model boundaries at the three mass ranges, we can
compute the total WDLF by summing the constituent low, intermediate and high
mass WDLFs by setting two of three models to \texttt{None} each time, and then
compared it to a WDLF computed using a single interpolated grid across
all three mass ranges.

In combination A, we use the pure Montreal stack. This naturally gives the
smoothest interpolated grid across the entire parameter space because of the
homogeneous choice of input physics and computation tools. Combination B makes
use of the most recent low, intermediate and high mass models from the
LPCODE\footnote{We use the smooth grid pre-interpolated by
their research group, as opposed to the individually published grids.}.
In combinations C and D, the low mass model is adopting that from the Montreal
group, while the intermediate and high mass models are from the BaSTI with and
without the inclusion of phase separation in the computation.

In B and C, the lump at the faintest end is due to the choice of ONe WDs for
$>1\,\msun$ WDs. They have different thermal properties to CO WDs so
they do not accumulate at the same bolometric magnitudes as the CO WDs when
cooling down.

\begin{table}
    \centering
    \begin{tabular}{cccc}
        Combination & Low Mass & Intermediate Mass & High Mass \\ \hline\hline
        A           & B20      & B20               & B20 \\
        B           & A09      & R10               & C19 \\
        C           & B20      & S10               & S10 \\
        D           & B20      & S10~(nps)         & S10~(nps) \\
    \end{tabular}
    \caption{Combination of the models used to compare the output WDLFs~(Refer
    to Table~\ref{tab:cooling_models} for the model names and properties).}
    \label{tab:cooling_model_combination}
\end{table}

\section{Other utilities}
\label{sec:utility}
There are two modules used in the background, \texttt{reddening} and
\texttt{util}; and one module, \texttt{plotter}, that could be useful for
basic inspection of the models. 

The \texttt{reddening} handles either choice of interpolation of the
extinctions, using a customised 2D spline interpolator if the tabulated values
from \citet{2011ApJ...737..103S} is used; or with the
\texttt{scipy.interpolate.RegularGridInterpolator} if the spectral type corrected
extinction values are chosen for dereddening. As mentioned in
Section~\ref{sec:gf21_comparison}, the extinction coefficient is tailored to the
temperature and the surface gravity of the target. Those coefficients are
obtained by convolving filter profiles with model spectra and convert the
total extinction into the extinction in each filter as a function of
temperature and surface gravity.

The other background module, \texttt{util}, supplies the customised 2D spline.
Figure~\ref{fig:cooling_tracks_default} is the default plot making use of the
\texttt{plotter} for displaying the Montreal atmosphere model. Similar
plots can be generated for inspecting the cooling models.

\section{Summary and Future Plan}
We provide a toolkit to allow easy access to synthetic photometry of WDs with
a number of formatters to handle the non-uniform model files and units from
various models. These models can be flexibly interpolated with two choices of
interpolators for immediate use with ease. Building on these interpolated
grids, this toolkit can perform photometric fitting of WDs with
or without a known distance, though the reliability is much higher if the distance is provided, and
it can also correct for reddening at each step of the iteration during the
minimisation/sampling.

We also make sure this grid provides a convenient and flexible
tool to construct WDLFs, allowing users to choose from a large number of
stellar evolution models and cooling models. We cover one set of DA and DB
atmosphere model, three initial mass functions, and three preset star formation
history models (constant, exponentially decay and burst). User-defined input
models can be supplied for each step of the WDLF computation (apart from
the cooling model), in order to allow a high level of flexibility when it
comes to more advanced usage.

In the future development of \textsc{WDPhotTools}, we intend to
\begin{enumerate}
    \item include new models that will become available;
    \item include synthetic photometry also provided in some of the cooling models;
    \item include means to use a user provided filter profile that can be convolved with the spectral energy distribution functions publicly available\footnote{\url{https://warwick.ac.uk/fac/sci/physics/research/astro/people/tremblay/modelgrids}};
    \item allow the use of different mass-radius relations;
    \item use user provided interstellar extinction correction;
    \item user supplied interpolator.
\end{enumerate}

In the coming decades, we shall expect higher quality distance estimates from
the future Gaia data releases, and deeper photometry couple with proper
motion measurements~(an alternative way to effectively select WD) with the
new generation facilities, for example, the Vera Rubin Observatory for the
southern sky, EUCLID and the Nancy Grace Roman Telescope for the study of the
halo WDs~\citep{2020ApJ...900..139F}. The \textsc{WDPhotTools}
will provide a user-friendly, convenient, and flexible tool for this
generation of astronomers to have an easier starting position and a more
gentle initial learning curve.

\section*{Acknowledgements}
The authors thank the anonymous referees for their comments that significantly
improve the readability of the articles.

MCL and MJG are supported by a European Research Council (ERC) grant under the European
Union’s Horizon 2020 research and innovation program (grant agreement number
833031).

MCL thanks Dr. N. Hambly and Dr. N. Rowell for their supports and guidance over
the years, as well as Dr. R. Smith, Prof \L{} Wyrzykowski, Dr. I. Arcavi,
and Prof. Dan Maoz for their supports to keep this research going over the
recent difficult situation around the globe.

MCL thanks Dr. P.-E. Tremblay and Dr. N. Gentile-Fusillo for cross-checking
some data.

This research has made use of the Spanish Virtual Observatory
(http://svo.cab.inta-csic.es) supported from the Spanish MICINN/FEDER through
grant AyA2017-84089.

This work has made use of data from the European Space Agency (ESA) mission
{\it Gaia} (\url{https://www.cosmos.esa.int/gaia}), processed by the {\it Gaia}
Data Processing and Analysis Consortium (DPAC,
\url{https://www.cosmos.esa.int/web/gaia/dpac/consortium}). Funding for the DPAC
has been provided by national institutions, in particular the institutions
participating in the {\it Gaia} Multilateral Agreement.

\section*{Data Availability}
The source code underlying this article are available on Github and Zenodo, at \url{https://github.com/cylammarco/WDPhotTools}, \url{https://doi.org/10.5281/zenodo.6570263}. The data and scripts for generating all the figures and analysis in this article, and the article itself, can be found at \url{https://github.com/cylammarco/WDPhotTools_article}.

The comparison data from \citet{2021MNRAS.508.3877G} can be found in the public domain at \url{https://cdsarc.cds.unistra.fr/viz-bin/cat/J/MNRAS/508/3877}.

The filter profile used in this work can be found in the public domain at \url{http://svo2.cab.inta-csic.es/theory/fps3/}.

The white dwarf theoretical spectra can be found in the public domain at \url{http://svo2.cab.inta-csic.es/theory/newov2/index.php?models=koester2}.

The Montreal models used in this work can be found in the public domain at \url{https://www.astro.umontreal.ca/~bergeron/CoolingModels/}.

The BASTI models used in this work can be found in the public domain at \url{http://albione.oa-teramo.inaf.it/}.

The LPCODE models used in this work can be found in the public domain at \url{http://evolgroup.fcaglp.unlp.edu.ar/TRACKS/tracks.html}.

The GENEVA stellar evolution models used in this work can be found in the public domain at \url{https://obswww.unige.ch/Research/evol/tables_grids2011/}.

The PARSEC stellar evolution models used in this work can be found in the public domain at \url{http://dx.doi.org/10.5281/zenodo.61584}.

The MIST stellar evolution models used in this work can be found in the public domain at \url{http://waps.cfa.harvard.edu/MIST/model_grids.html}.



\bibliographystyle{rasti}
\bibliography{WDPhotTools} 

\begin{thebibliography}{}
\makeatletter
\relax
\def\mn@urlcharsother{\let\do\@makeother \do\$\do\&\do\#\do\^\do\_\do\%\do\~}
\def\mn@doi{\begingroup\mn@urlcharsother \@ifnextchar [ {\mn@doi@}
  {\mn@doi@[]}}
\def\mn@doi@[#1]#2{\def\@tempa{#1}\ifx\@tempa\@empty \href
  {http://dx.doi.org/#2} {doi:#2}\else \href {http://dx.doi.org/#2} {#1}\fi
  \endgroup}
\def\mn@eprint#1#2{\mn@eprint@#1:#2::\@nil}
\def\mn@eprint@arXiv#1{\href {http://arxiv.org/abs/#1} {{\tt arXiv:#1}}}
\def\mn@eprint@dblp#1{\href {http://dblp.uni-trier.de/rec/bibtex/#1.xml}
  {dblp:#1}}
\def\mn@eprint@#1:#2:#3:#4\@nil{\def\@tempa {#1}\def\@tempb {#2}\def\@tempc
  {#3}\ifx \@tempc \@empty \let \@tempc \@tempb \let \@tempb \@tempa \fi \ifx
  \@tempb \@empty \def\@tempb {arXiv}\fi \@ifundefined
  {mn@eprint@\@tempb}{\@tempb:\@tempc}{\expandafter \expandafter \csname
  mn@eprint@\@tempb\endcsname \expandafter{\@tempc}}}

\bibitem[\protect\citeauthoryear{{Althaus}, {Garc{\'\i}a-Berro}, {Isern},
  {C{\'o}rsico}  \& {Rohrmann}}{{Althaus} et~al.}{2007}]{2007A&A...465..249A}
{Althaus} L.~G.,  {Garc{\'\i}a-Berro} E.,  {Isern} J.,  {C{\'o}rsico} A.~H.,
  {Rohrmann} R.~D.,  2007, \mn@doi [\aap] {10.1051/0004-6361:20066059}, \href
  {https://ui.adsabs.harvard.edu/abs/2007A&A...465..249A} {465, 249}

\bibitem[\protect\citeauthoryear{{Althaus}, {Panei}, {Miller Bertolami},
  {Garc{\'\i}a-Berro}, {C{\'o}rsico}, {Romero}, {Kepler}  \&
  {Rohrmann}}{{Althaus} et~al.}{2009}]{2009ApJ...704.1605A}
{Althaus} L.~G.,  {Panei} J.~A.,  {Miller Bertolami} M.~M.,
  {Garc{\'\i}a-Berro} E.,  {C{\'o}rsico} A.~H.,  {Romero} A.~D.,  {Kepler}
  S.~O.,   {Rohrmann} R.~D.,  2009, \mn@doi [\apj]
  {10.1088/0004-637X/704/2/1605}, \href
  {https://ui.adsabs.harvard.edu/abs/2009ApJ...704.1605A} {704, 1605}

\bibitem[\protect\citeauthoryear{{Althaus}, {C{\'o}rsico}, {Isern}  \&
  {Garc{\'\i}a-Berro}}{{Althaus} et~al.}{2010a}]{2010A&ARv..18..471A}
{Althaus} L.~G.,  {C{\'o}rsico} A.~H.,  {Isern} J.,   {Garc{\'\i}a-Berro} E.,
  2010a, \mn@doi [\aapr] {10.1007/s00159-010-0033-1}, \href
  {https://ui.adsabs.harvard.edu/abs/2010A&ARv..18..471A} {18, 471}

\bibitem[\protect\citeauthoryear{{Althaus}, {Garc{\'\i}a-Berro}, {Renedo},
  {Isern}, {C{\'o}rsico}  \& {Rohrmann}}{{Althaus}
  et~al.}{2010b}]{2010ApJ...719..612A}
{Althaus} L.~G.,  {Garc{\'\i}a-Berro} E.,  {Renedo} I.,  {Isern} J.,
  {C{\'o}rsico} A.~H.,   {Rohrmann} R.~D.,  2010b, \mn@doi [\apj]
  {10.1088/0004-637X/719/1/612}, \href
  {https://ui.adsabs.harvard.edu/abs/2010ApJ...719..612A} {719, 612}

\bibitem[\protect\citeauthoryear{{Althaus}, {Miller Bertolami}  \&
  {C{\'o}rsico}}{{Althaus} et~al.}{2013}]{2013A&A...557A..19A}
{Althaus} L.~G.,  {Miller Bertolami} M.~M.,   {C{\'o}rsico} A.~H.,  2013,
  \mn@doi [\aap] {10.1051/0004-6361/201321868}, \href
  {https://ui.adsabs.harvard.edu/abs/2013A&A...557A..19A} {557, A19}

\bibitem[\protect\citeauthoryear{{Althaus}, {Camisassa}, {Miller Bertolami},
  {C{\'o}rsico}  \& {Garc{\'\i}a-Berro}}{{Althaus}
  et~al.}{2015}]{2015A&A...576A...9A}
{Althaus} L.~G.,  {Camisassa} M.~E.,  {Miller Bertolami} M.~M.,  {C{\'o}rsico}
  A.~H.,   {Garc{\'\i}a-Berro} E.,  2015, \mn@doi [\aap]
  {10.1051/0004-6361/201424922}, \href
  {https://ui.adsabs.harvard.edu/abs/2015A&A...576A...9A} {576, A9}

\bibitem[\protect\citeauthoryear{{Althaus}, {De Ger{\'o}nimo}, {C{\'o}rsico},
  {Torres}  \& {Garc{\'\i}a-Berro}}{{Althaus}
  et~al.}{2017}]{2017A&A...597A..67A}
{Althaus} L.~G.,  {De Ger{\'o}nimo} F.,  {C{\'o}rsico} A.,  {Torres} S.,
  {Garc{\'\i}a-Berro} E.,  2017, \mn@doi [\aap] {10.1051/0004-6361/201629909},
  \href {https://ui.adsabs.harvard.edu/abs/2017A&A...597A..67A} {597, A67}

\bibitem[\protect\citeauthoryear{{Andrews}, {Ag{\"u}eros}, {Gianninas},
  {Kilic}, {Dhital}  \& {Anderson}}{{Andrews}
  et~al.}{2015}]{2015ApJ...815...63A}
{Andrews} J.~J.,  {Ag{\"u}eros} M.~A.,  {Gianninas} A.,  {Kilic} M.,  {Dhital}
  S.,   {Anderson} S.~F.,  2015, \mn@doi [\apj] {10.1088/0004-637X/815/1/63},
  \href {https://ui.adsabs.harvard.edu/abs/2015ApJ...815...63A} {815, 63}

\bibitem[\protect\citeauthoryear{{Bailer-Jones}, {Rybizki}, {Fouesneau},
  {Demleitner}  \& {Andrae}}{{Bailer-Jones} et~al.}{2021}]{2021AJ....161..147B}
{Bailer-Jones} C.~A.~L.,  {Rybizki} J.,  {Fouesneau} M.,  {Demleitner} M.,
  {Andrae} R.,  2021, \mn@doi [\aj] {10.3847/1538-3881/abd806}, \href
  {https://ui.adsabs.harvard.edu/abs/2021AJ....161..147B} {161, 147}

\bibitem[\protect\citeauthoryear{{Barbary}}{{Barbary}}{2016}]{2016zndo....804967B}
{Barbary} K.,  2016, {Extinction V0.3.0}, Zenodo,
  \mn@doi{10.5281/zenodo.804967}

\bibitem[\protect\citeauthoryear{{Barrientos} \& {Chanam{\'e}}}{{Barrientos} \&
  {Chanam{\'e}}}{2021}]{2021ApJ...923..181B}
{Barrientos} M.,  {Chanam{\'e}} J.,  2021, \mn@doi [\apj]
  {10.3847/1538-4357/ac2f49}, \href
  {https://ui.adsabs.harvard.edu/abs/2021ApJ...923..181B} {923, 181}

\bibitem[\protect\citeauthoryear{{B{\'e}dard}, {Bergeron}, {Brassard}  \&
  {Fontaine}}{{B{\'e}dard} et~al.}{2020}]{2020ApJ...901...93B}
{B{\'e}dard} A.,  {Bergeron} P.,  {Brassard} P.,   {Fontaine} G.,  2020,
  \mn@doi [\apj] {10.3847/1538-4357/abafbe}, \href
  {https://ui.adsabs.harvard.edu/abs/2020ApJ...901...93B} {901, 93}

\bibitem[\protect\citeauthoryear{{Bedin}, {King}, {Anderson}, {Piotto},
  {Salaris}, {Cassisi}  \& {Serenelli}}{{Bedin}
  et~al.}{2008}]{2008ApJ...678.1279B}
{Bedin} L.~R.,  {King} I.~R.,  {Anderson} J.,  {Piotto} G.,  {Salaris} M.,
  {Cassisi} S.,   {Serenelli} A.,  2008, \mn@doi [\apj] {10.1086/529370}, \href
  {https://ui.adsabs.harvard.edu/abs/2008ApJ...678.1279B} {678, 1279}

\bibitem[\protect\citeauthoryear{{Bergeron}, {Wesemael}  \&
  {Beauchamp}}{{Bergeron} et~al.}{1995}]{1995PASP..107.1047B}
{Bergeron} P.,  {Wesemael} F.,   {Beauchamp} A.,  1995, \mn@doi [\pasp]
  {10.1086/133661}, \href
  {https://ui.adsabs.harvard.edu/abs/1995PASP..107.1047B} {107, 1047}

\bibitem[\protect\citeauthoryear{{Bergeron} et~al.,}{{Bergeron}
  et~al.}{2011}]{2011ApJ...737...28B}
{Bergeron} P.,  et~al., 2011, \mn@doi [\apj] {10.1088/0004-637X/737/1/28},
  \href {https://ui.adsabs.harvard.edu/abs/2011ApJ...737...28B} {737, 28}

\bibitem[\protect\citeauthoryear{{Bergeron}, {Dufour}, {Fontaine}, {Coutu},
  {Blouin}, {Genest-Beaulieu}, {B{\'e}dard}  \& {Rolland}}{{Bergeron}
  et~al.}{2019}]{2019ApJ...876...67B}
{Bergeron} P.,  {Dufour} P.,  {Fontaine} G.,  {Coutu} S.,  {Blouin} S.,
  {Genest-Beaulieu} C.,  {B{\'e}dard} A.,   {Rolland} B.,  2019, \mn@doi [\apj]
  {10.3847/1538-4357/ab153a}, \href
  {https://ui.adsabs.harvard.edu/abs/2019ApJ...876...67B} {876, 67}

\bibitem[\protect\citeauthoryear{{Blouin}, {Kowalski}  \& {Dufour}}{{Blouin}
  et~al.}{2017}]{2017ApJ...848...36B}
{Blouin} S.,  {Kowalski} P.~M.,   {Dufour} P.,  2017, \mn@doi [\apj]
  {10.3847/1538-4357/aa8ad6}, \href
  {https://ui.adsabs.harvard.edu/abs/2017ApJ...848...36B} {848, 36}

\bibitem[\protect\citeauthoryear{{Blouin}, {Dufour}  \& {Allard}}{{Blouin}
  et~al.}{2018}]{2018ApJ...863..184B}
{Blouin} S.,  {Dufour} P.,   {Allard} N.~F.,  2018, \mn@doi [\apj]
  {10.3847/1538-4357/aad4a9}, \href
  {https://ui.adsabs.harvard.edu/abs/2018ApJ...863..184B} {863, 184}

\bibitem[\protect\citeauthoryear{{Bressan}, {Marigo}, {Girardi}, {Salasnich},
  {Dal Cero}, {Rubele}  \& {Nanni}}{{Bressan}
  et~al.}{2012}]{2012MNRAS.427..127B}
{Bressan} A.,  {Marigo} P.,  {Girardi} L.,  {Salasnich} B.,  {Dal Cero} C.,
  {Rubele} S.,   {Nanni} A.,  2012, \mn@doi [\mnras]
  {10.1111/j.1365-2966.2012.21948.x}, \href
  {https://ui.adsabs.harvard.edu/abs/2012MNRAS.427..127B} {427, 127}

\bibitem[\protect\citeauthoryear{{Bressan}, {Marigo}, {Girardi}, {Nanni}  \&
  {Rubele}}{{Bressan} et~al.}{2013}]{2013EPJWC..4303001B}
{Bressan} A.,  {Marigo} P.,  {Girardi} L.,  {Nanni} A.,   {Rubele} S.,  2013,
  in European Physical Journal Web of Conferences. p. 03001 (\mn@eprint {arXiv}
  {1301.7687}), \mn@doi{10.1051/epjconf/20134303001}

\bibitem[\protect\citeauthoryear{{Camisassa}, {Althaus}, {C{\'o}rsico},
  {Vinyoles}, {Serenelli}, {Isern}, {Miller Bertolami}  \&
  {Garc{\'\i}a{\textendash}Berro}}{{Camisassa}
  et~al.}{2016}]{2016ApJ...823..158C}
{Camisassa} M.~E.,  {Althaus} L.~G.,  {C{\'o}rsico} A.~H.,  {Vinyoles} N.,
  {Serenelli} A.~M.,  {Isern} J.,  {Miller Bertolami} M.~M.,
  {Garc{\'\i}a{\textendash}Berro} E.,  2016, \mn@doi [\apj]
  {10.3847/0004-637X/823/2/158}, \href
  {https://ui.adsabs.harvard.edu/abs/2016ApJ...823..158C} {823, 158}

\bibitem[\protect\citeauthoryear{{Camisassa}, {Althaus}, {Rohrmann},
  {Garc{\'\i}a-Berro}, {Torres}, {C{\'o}rsico}  \& {Wachlin}}{{Camisassa}
  et~al.}{2017}]{2017ApJ...839...11C}
{Camisassa} M.~E.,  {Althaus} L.~G.,  {Rohrmann} R.~D.,  {Garc{\'\i}a-Berro}
  E.,  {Torres} S.,  {C{\'o}rsico} A.~H.,   {Wachlin} F.~C.,  2017, \mn@doi
  [\apj] {10.3847/1538-4357/aa6797}, \href
  {https://ui.adsabs.harvard.edu/abs/2017ApJ...839...11C} {839, 11}

\bibitem[\protect\citeauthoryear{{Camisassa} et~al.,}{{Camisassa}
  et~al.}{2019}]{2019A&A...625A..87C}
{Camisassa} M.~E.,  et~al., 2019, \mn@doi [\aap] {10.1051/0004-6361/201833822},
  \href {https://ui.adsabs.harvard.edu/abs/2019A&A...625A..87C} {625, A87}

\bibitem[\protect\citeauthoryear{{Catal{\'a}n}}{{Catal{\'a}n}}{2015}]{2015ASPC..493..325C}
{Catal{\'a}n} S.,  2015, in {Dufour} P.,  {Bergeron} P.,   {Fontaine} G.,  eds,
   Astronomical Society of the Pacific Conference Series Vol. 493, 19th
  European Workshop on White Dwarfs. p.~325

\bibitem[\protect\citeauthoryear{{Catal{\'a}n}, {Isern}, {Garc{\'\i}a-Berro}
  \& {Ribas}}{{Catal{\'a}n} et~al.}{2008a}]{2008MNRAS.387.1693C}
{Catal{\'a}n} S.,  {Isern} J.,  {Garc{\'\i}a-Berro} E.,   {Ribas} I.,  2008a,
  \mn@doi [\mnras] {10.1111/j.1365-2966.2008.13356.x}, \href
  {https://ui.adsabs.harvard.edu/abs/2008MNRAS.387.1693C} {387, 1693}

\bibitem[\protect\citeauthoryear{{Catal{\'a}n}, {Isern}, {Garc{\'\i}a-Berro},
  {Ribas}, {Allende Prieto}  \& {Bonanos}}{{Catal{\'a}n}
  et~al.}{2008b}]{2008A&A...477..213C}
{Catal{\'a}n} S.,  {Isern} J.,  {Garc{\'\i}a-Berro} E.,  {Ribas} I.,  {Allende
  Prieto} C.,   {Bonanos} A.~Z.,  2008b, \mn@doi [\aap]
  {10.1051/0004-6361:20078111}, \href
  {https://ui.adsabs.harvard.edu/abs/2008A&A...477..213C} {477, 213}

\bibitem[\protect\citeauthoryear{{Chabrier}}{{Chabrier}}{2003}]{2003PASP..115..763C}
{Chabrier} G.,  2003, \mn@doi [\pasp] {10.1086/376392}, \href
  {https://ui.adsabs.harvard.edu/abs/2003PASP..115..763C} {115, 763}

\bibitem[\protect\citeauthoryear{{Chandra}, {Hwang}, {Zakamska}  \&
  {Budav{\'a}ri}}{{Chandra} et~al.}{2020}]{2020MNRAS.497.2688C}
{Chandra} V.,  {Hwang} H.-C.,  {Zakamska} N.~L.,   {Budav{\'a}ri} T.,  2020,
  \mn@doi [\mnras] {10.1093/mnras/staa2165}, \href
  {https://ui.adsabs.harvard.edu/abs/2020MNRAS.497.2688C} {497, 2688}

\bibitem[\protect\citeauthoryear{{Chen}, {Girardi}, {Bressan}, {Marigo},
  {Barbieri}  \& {Kong}}{{Chen} et~al.}{2014}]{2014MNRAS.444.2525C}
{Chen} Y.,  {Girardi} L.,  {Bressan} A.,  {Marigo} P.,  {Barbieri} M.,   {Kong}
  X.,  2014, \mn@doi [\mnras] {10.1093/mnras/stu1605}, \href
  {https://ui.adsabs.harvard.edu/abs/2014MNRAS.444.2525C} {444, 2525}

\bibitem[\protect\citeauthoryear{{Choi}, {Dotter}, {Conroy}, {Cantiello},
  {Paxton}  \& {Johnson}}{{Choi} et~al.}{2016}]{2016ApJ...823..102C}
{Choi} J.,  {Dotter} A.,  {Conroy} C.,  {Cantiello} M.,  {Paxton} B.,
  {Johnson} B.~D.,  2016, \mn@doi [\apj] {10.3847/0004-637X/823/2/102}, \href
  {https://ui.adsabs.harvard.edu/abs/2016ApJ...823..102C} {823, 102}

\bibitem[\protect\citeauthoryear{{Cummings}, {Kalirai}, {Tremblay}  \&
  {Ramirez-Ruiz}}{{Cummings} et~al.}{2016}]{2016ApJ...818...84C}
{Cummings} J.~D.,  {Kalirai} J.~S.,  {Tremblay} P.~E.,   {Ramirez-Ruiz} E.,
  2016, \mn@doi [\apj] {10.3847/0004-637X/818/1/84}, \href
  {https://ui.adsabs.harvard.edu/abs/2016ApJ...818...84C} {818, 84}

\bibitem[\protect\citeauthoryear{{Cummings}, {Kalirai}, {Tremblay},
  {Ramirez-Ruiz}  \& {Choi}}{{Cummings} et~al.}{2018}]{2018ApJ...866...21C}
{Cummings} J.~D.,  {Kalirai} J.~S.,  {Tremblay} P.~E.,  {Ramirez-Ruiz} E.,
  {Choi} J.,  2018, \mn@doi [\apj] {10.3847/1538-4357/aadfd6}, \href
  {https://ui.adsabs.harvard.edu/abs/2018ApJ...866...21C} {866, 21}

\bibitem[\protect\citeauthoryear{{D'Antona} \& {Mazzitelli}}{{D'Antona} \&
  {Mazzitelli}}{1989}]{1989ApJ...347..934D}
{D'Antona} F.,  {Mazzitelli} I.,  1989, \mn@doi [\apj] {10.1086/168185}, \href
  {https://ui.adsabs.harvard.edu/abs/1989ApJ...347..934D} {347, 934}

\bibitem[\protect\citeauthoryear{{Deloye} \& {Bildsten}}{{Deloye} \&
  {Bildsten}}{2002}]{2002ApJ...580.1077D}
{Deloye} C.~J.,  {Bildsten} L.,  2002, \mn@doi [\apj] {10.1086/343800}, \href
  {https://ui.adsabs.harvard.edu/abs/2002ApJ...580.1077D} {580, 1077}

\bibitem[\protect\citeauthoryear{{Dotter}}{{Dotter}}{2016}]{2016ApJS..222....8D}
{Dotter} A.,  2016, \mn@doi [\apjs] {10.3847/0067-0049/222/1/8}, \href
  {https://ui.adsabs.harvard.edu/abs/2016ApJS..222....8D} {222, 8}

\bibitem[\protect\citeauthoryear{{Ekstr{\"o}m} et~al.,}{{Ekstr{\"o}m}
  et~al.}{2012}]{2012A&A...537A.146E}
{Ekstr{\"o}m} S.,  et~al., 2012, \mn@doi [\aap] {10.1051/0004-6361/201117751},
  \href {https://ui.adsabs.harvard.edu/abs/2012A&A...537A.146E} {537, A146}

\bibitem[\protect\citeauthoryear{{El-Badry}, {Rix}  \& {Weisz}}{{El-Badry}
  et~al.}{2018}]{2018ApJ...860L..17E}
{El-Badry} K.,  {Rix} H.-W.,   {Weisz} D.~R.,  2018, \mn@doi [\apjl]
  {10.3847/2041-8213/aaca9c}, \href
  {https://ui.adsabs.harvard.edu/abs/2018ApJ...860L..17E} {860, L17}

\bibitem[\protect\citeauthoryear{{Fantin}, {C{\^o}t{\'e}}  \&
  {McConnachie}}{{Fantin} et~al.}{2020}]{2020ApJ...900..139F}
{Fantin} N.~J.,  {C{\^o}t{\'e}} P.,   {McConnachie} A.~W.,  2020, \mn@doi
  [\apj] {10.3847/1538-4357/aba270}, \href
  {https://ui.adsabs.harvard.edu/abs/2020ApJ...900..139F} {900, 139}

\bibitem[\protect\citeauthoryear{{Fitzpatrick}}{{Fitzpatrick}}{1999}]{1999PASP..111...63F}
{Fitzpatrick} E.~L.,  1999, \mn@doi [\pasp] {10.1086/316293}, \href
  {https://ui.adsabs.harvard.edu/abs/1999PASP..111...63F} {111, 63}

\bibitem[\protect\citeauthoryear{{Fontaine}, {Brassard}  \&
  {Bergeron}}{{Fontaine} et~al.}{2001}]{2001PASP..113..409F}
{Fontaine} G.,  {Brassard} P.,   {Bergeron} P.,  2001, \mn@doi [\pasp]
  {10.1086/319535}, \href
  {https://ui.adsabs.harvard.edu/abs/2001PASP..113..409F} {113, 409}

\bibitem[\protect\citeauthoryear{{Foreman-Mackey}, {Hogg}, {Lang}  \&
  {Goodman}}{{Foreman-Mackey} et~al.}{2013}]{2013PASP..125..306F}
{Foreman-Mackey} D.,  {Hogg} D.~W.,  {Lang} D.,   {Goodman} J.,  2013, \mn@doi
  [\pasp] {10.1086/670067}, \href
  {https://ui.adsabs.harvard.edu/abs/2013PASP..125..306F} {125, 306}

\bibitem[\protect\citeauthoryear{{Gaia Collaboration} et~al.,}{{Gaia
  Collaboration} et~al.}{2021a}]{2021A&A...649A...1G}
{Gaia Collaboration} et~al., 2021a, \mn@doi [\aap]
  {10.1051/0004-6361/202039657}, \href
  {https://ui.adsabs.harvard.edu/abs/2021A&A...649A...1G} {649, A1}

\bibitem[\protect\citeauthoryear{{Gaia Collaboration} et~al.,}{{Gaia
  Collaboration} et~al.}{2021b}]{2021A&A...649A...6G}
{Gaia Collaboration} et~al., 2021b, \mn@doi [\aap]
  {10.1051/0004-6361/202039498}, \href
  {https://ui.adsabs.harvard.edu/abs/2021A&A...649A...6G} {649, A6}

\bibitem[\protect\citeauthoryear{{Garc{\'\i}a-Berro}, {Althaus}, {C{\'o}rsico}
  \& {Isern}}{{Garc{\'\i}a-Berro} et~al.}{2008}]{2008ApJ...677..473G}
{Garc{\'\i}a-Berro} E.,  {Althaus} L.~G.,  {C{\'o}rsico} A.~H.,   {Isern} J.,
  2008, \mn@doi [\apj] {10.1086/527536}, \href
  {https://ui.adsabs.harvard.edu/abs/2008ApJ...677..473G} {677, 473}

\bibitem[\protect\citeauthoryear{{Garc{\'\i}a-Berro}
  et~al.,}{{Garc{\'\i}a-Berro} et~al.}{2010}]{2010Natur.465..194G}
{Garc{\'\i}a-Berro} E.,  et~al., 2010, \mn@doi [\nat] {10.1038/nature09045},
  \href {https://ui.adsabs.harvard.edu/abs/2010Natur.465..194G} {465, 194}

\bibitem[\protect\citeauthoryear{{Genest-Beaulieu} \&
  {Bergeron}}{{Genest-Beaulieu} \& {Bergeron}}{2019a}]{2019ApJ...871..169G}
{Genest-Beaulieu} C.,  {Bergeron} P.,  2019a, \mn@doi [\apj]
  {10.3847/1538-4357/aafac6}, \href
  {https://ui.adsabs.harvard.edu/abs/2019ApJ...871..169G} {871, 169}

\bibitem[\protect\citeauthoryear{{Genest-Beaulieu} \&
  {Bergeron}}{{Genest-Beaulieu} \& {Bergeron}}{2019b}]{2019ApJ...882..106G}
{Genest-Beaulieu} C.,  {Bergeron} P.,  2019b, \mn@doi [\apj]
  {10.3847/1538-4357/ab379e}, \href
  {https://ui.adsabs.harvard.edu/abs/2019ApJ...882..106G} {882, 106}

\bibitem[\protect\citeauthoryear{{Gentile Fusillo} et~al.,}{{Gentile Fusillo}
  et~al.}{2021}]{2021MNRAS.508.3877G}
{Gentile Fusillo} N.~P.,  et~al., 2021, \mn@doi [\mnras]
  {10.1093/mnras/stab2672}, \href
  {https://ui.adsabs.harvard.edu/abs/2021MNRAS.508.3877G} {508, 3877}

\bibitem[\protect\citeauthoryear{{Georgy}, {Ekstr{\"o}m}, {Granada}, {Meynet},
  {Mowlavi}, {Eggenberger}  \& {Maeder}}{{Georgy}
  et~al.}{2013a}]{2013A&A...553A..24G}
{Georgy} C.,  {Ekstr{\"o}m} S.,  {Granada} A.,  {Meynet} G.,  {Mowlavi} N.,
  {Eggenberger} P.,   {Maeder} A.,  2013a, \mn@doi [\aap]
  {10.1051/0004-6361/201220558}, \href
  {https://ui.adsabs.harvard.edu/abs/2013A&A...553A..24G} {553, A24}

\bibitem[\protect\citeauthoryear{{Georgy} et~al.,}{{Georgy}
  et~al.}{2013b}]{2013A&A...558A.103G}
{Georgy} C.,  et~al., 2013b, \mn@doi [\aap] {10.1051/0004-6361/201322178},
  \href {https://ui.adsabs.harvard.edu/abs/2013A&A...558A.103G} {558, A103}

\bibitem[\protect\citeauthoryear{{Giammichele}, {Bergeron}  \&
  {Dufour}}{{Giammichele} et~al.}{2012}]{2012ApJS..199...29G}
{Giammichele} N.,  {Bergeron} P.,   {Dufour} P.,  2012, \mn@doi [\apjs]
  {10.1088/0067-0049/199/2/29}, \href
  {https://ui.adsabs.harvard.edu/abs/2012ApJS..199...29G} {199, 29}

\bibitem[\protect\citeauthoryear{{Haft}, {Raffelt}  \& {Weiss}}{{Haft}
  et~al.}{1994}]{1994ApJ...425..222H}
{Haft} M.,  {Raffelt} G.,   {Weiss} A.,  1994, \mn@doi [\apj] {10.1086/173978},
  \href {https://ui.adsabs.harvard.edu/abs/1994ApJ...425..222H} {425, 222}

\bibitem[\protect\citeauthoryear{{Harris} et~al.,}{{Harris}
  et~al.}{2006}]{2006AJ....131..571H}
{Harris} H.~C.,  et~al., 2006, \mn@doi [\aj] {10.1086/497966}, \href
  {https://ui.adsabs.harvard.edu/abs/2006AJ....131..571H} {131, 571}

\bibitem[\protect\citeauthoryear{{Harris} et~al.,}{{Harris}
  et~al.}{2020}]{2020Natur.585..357H}
{Harris} C.~R.,  et~al., 2020, \mn@doi [\nat] {10.1038/s41586-020-2649-2},
  \href {https://ui.adsabs.harvard.edu/abs/2020Natur.585..357H} {585, 357}

\bibitem[\protect\citeauthoryear{{Heyl}, {Caiazzo}  \& {Richer}}{{Heyl}
  et~al.}{2022}]{2022ApJ...926..132H}
{Heyl} J.,  {Caiazzo} I.,   {Richer} H.~B.,  2022, \mn@doi [\apj]
  {10.3847/1538-4357/ac45fc}, \href
  {https://ui.adsabs.harvard.edu/abs/2022ApJ...926..132H} {926, 132}

\bibitem[\protect\citeauthoryear{{Holberg} \& {Bergeron}}{{Holberg} \&
  {Bergeron}}{2006}]{2006AJ....132.1221H}
{Holberg} J.~B.,  {Bergeron} P.,  2006, \mn@doi [\aj] {10.1086/505938}, \href
  {https://ui.adsabs.harvard.edu/abs/2006AJ....132.1221H} {132, 1221}

\bibitem[\protect\citeauthoryear{{Iben} \& {Tutukov}}{{Iben} \&
  {Tutukov}}{1984}]{1984ApJ...282..615I}
{Iben} I. J.,  {Tutukov} A.~V.,  1984, \mn@doi [\apj] {10.1086/162241}, \href
  {https://ui.adsabs.harvard.edu/abs/1984ApJ...282..615I} {282, 615}

\bibitem[\protect\citeauthoryear{{Itoh}, {Hayashi}, {Nishikawa}  \&
  {Kohyama}}{{Itoh} et~al.}{1996}]{1996ApJS..102..411I}
{Itoh} N.,  {Hayashi} H.,  {Nishikawa} A.,   {Kohyama} Y.,  1996, \mn@doi
  [\apjs] {10.1086/192264}, \href
  {https://ui.adsabs.harvard.edu/abs/1996ApJS..102..411I} {102, 411}

\bibitem[\protect\citeauthoryear{{Kalirai}, {Saul Davis}, {Richer}, {Bergeron},
  {Catelan}, {Hansen}  \& {Rich}}{{Kalirai} et~al.}{2009}]{2009ApJ...705..408K}
{Kalirai} J.~S.,  {Saul Davis} D.,  {Richer} H.~B.,  {Bergeron} P.,  {Catelan}
  M.,  {Hansen} B. M.~S.,   {Rich} R.~M.,  2009, \mn@doi [\apj]
  {10.1088/0004-637X/705/1/408}, \href
  {https://ui.adsabs.harvard.edu/abs/2009ApJ...705..408K} {705, 408}

\bibitem[\protect\citeauthoryear{{Kepler}, {Koester}, {Pelisoli}, {Romero}  \&
  {Ourique}}{{Kepler} et~al.}{2021}]{2021MNRAS.507.4646K}
{Kepler} S.~O.,  {Koester} D.,  {Pelisoli} I.,  {Romero} A.~D.,   {Ourique} G.,
   2021, \mn@doi [\mnras] {10.1093/mnras/stab2411}, \href
  {https://ui.adsabs.harvard.edu/abs/2021MNRAS.507.4646K} {507, 4646}

\bibitem[\protect\citeauthoryear{{Kilic}, {Stanek}  \& {Pinsonneault}}{{Kilic}
  et~al.}{2007}]{2007ApJ...671..761K}
{Kilic} M.,  {Stanek} K.~Z.,   {Pinsonneault} M.~H.,  2007, \mn@doi [\apj]
  {10.1086/522228}, \href
  {https://ui.adsabs.harvard.edu/abs/2007ApJ...671..761K} {671, 761}

\bibitem[\protect\citeauthoryear{{Kippenhahn}, {Weigert}  \&
  {Weiss}}{{Kippenhahn} et~al.}{2013}]{2013sse..book.....K}
{Kippenhahn} R.,  {Weigert} A.,   {Weiss} A.,  2013, {Stellar Structure and
  Evolution}, \mn@doi{10.1007/978-3-642-30304-3.
}

\bibitem[\protect\citeauthoryear{{Knox}, {Hawkins}  \& {Hambly}}{{Knox}
  et~al.}{1999}]{1999MNRAS.306..736K}
{Knox} R.~A.,  {Hawkins} M.~R.~S.,   {Hambly} N.~C.,  1999, \mn@doi [\mnras]
  {10.1046/j.1365-8711.1999.02625.x}, \href
  {https://ui.adsabs.harvard.edu/abs/1999MNRAS.306..736K} {306, 736}

\bibitem[\protect\citeauthoryear{{Koester}}{{Koester}}{2010}]{2010MmSAI..81..921K}
{Koester} D.,  2010, \memsai, \href
  {https://ui.adsabs.harvard.edu/abs/2010MmSAI..81..921K} {81, 921}

\bibitem[\protect\citeauthoryear{{Kowalski} \& {Saumon}}{{Kowalski} \&
  {Saumon}}{2006}]{2006ApJ...651L.137K}
{Kowalski} P.~M.,  {Saumon} D.,  2006, \mn@doi [\apjl] {10.1086/509723}, \href
  {https://ui.adsabs.harvard.edu/abs/2006ApJ...651L.137K} {651, L137}

\bibitem[\protect\citeauthoryear{{Kroupa}}{{Kroupa}}{2001}]{2001MNRAS.322..231K}
{Kroupa} P.,  2001, \mn@doi [\mnras] {10.1046/j.1365-8711.2001.04022.x}, \href
  {https://ui.adsabs.harvard.edu/abs/2001MNRAS.322..231K} {322, 231}

\bibitem[\protect\citeauthoryear{{Lam}}{{Lam}}{2017}]{2017ASPC..509...25L}
{Lam} M.~C.,  2017, in {Tremblay} P.~E.,  {Gaensicke} B.,   {Marsh} T.,  eds,
  Astronomical Society of the Pacific Conference Series Vol. 509, 20th European
  White Dwarf Workshop. p.~25 (\mn@eprint {arXiv} {1702.02187})

\bibitem[\protect\citeauthoryear{{Lam} et~al.,}{{Lam}
  et~al.}{2019}]{2019MNRAS.482..715L}
{Lam} M.~C.,  et~al., 2019, \mn@doi [\mnras] {10.1093/mnras/sty2710}, \href
  {https://ui.adsabs.harvard.edu/abs/2019MNRAS.482..715L} {482, 715}

\bibitem[\protect\citeauthoryear{{Lam}, {Hambly}, {Lodieu}, {Blouin}, {Harvey},
  {Smith}, {G{\'a}lvez-Ortiz}  \& {Zhang}}{{Lam}
  et~al.}{2020}]{2020MNRAS.493.6001L}
{Lam} M.~C.,  {Hambly} N.~C.,  {Lodieu} N.,  {Blouin} S.,  {Harvey} E.~J.,
  {Smith} R.~J.,  {G{\'a}lvez-Ortiz} M.~C.,   {Zhang} Z.~H.,  2020, \mn@doi
  [\mnras] {10.1093/mnras/staa584}, \href
  {https://ui.adsabs.harvard.edu/abs/2020MNRAS.493.6001L} {493, 6001}

\bibitem[\protect\citeauthoryear{{Lauffer}, {Romero}  \& {Kepler}}{{Lauffer}
  et~al.}{2018}]{2018MNRAS.480.1547L}
{Lauffer} G.~R.,  {Romero} A.~D.,   {Kepler} S.~O.,  2018, \mn@doi [\mnras]
  {10.1093/mnras/sty1925}, \href
  {https://ui.adsabs.harvard.edu/abs/2018MNRAS.480.1547L} {480, 1547}

\bibitem[\protect\citeauthoryear{{Leggett}, {Ruiz}  \& {Bergeron}}{{Leggett}
  et~al.}{1998}]{1998ApJ...497..294L}
{Leggett} S.~K.,  {Ruiz} M.~T.,   {Bergeron} P.,  1998, \mn@doi [\apj]
  {10.1086/305463}, \href
  {https://ui.adsabs.harvard.edu/abs/1998ApJ...497..294L} {497, 294}

\bibitem[\protect\citeauthoryear{{Liebert}, {Dahn}, {Gresham}  \&
  {Strittmatter}}{{Liebert} et~al.}{1979}]{1979ApJ...233..226L}
{Liebert} J.,  {Dahn} C.~C.,  {Gresham} M.,   {Strittmatter} P.~A.,  1979,
  \mn@doi [\apj] {10.1086/157384}, \href
  {https://ui.adsabs.harvard.edu/abs/1979ApJ...233..226L} {233, 226}

\bibitem[\protect\citeauthoryear{{Liebert}, {Dahn}  \& {Monet}}{{Liebert}
  et~al.}{1988}]{1988ApJ...332..891L}
{Liebert} J.,  {Dahn} C.~C.,   {Monet} D.~G.,  1988, \mn@doi [\apj]
  {10.1086/166699}, \href
  {https://ui.adsabs.harvard.edu/abs/1988ApJ...332..891L} {332, 891}

\bibitem[\protect\citeauthoryear{{Liebert}, {Dahn}  \& {Monet}}{{Liebert}
  et~al.}{1989}]{1989LNP...328...15L}
{Liebert} J.,  {Dahn} C.~C.,   {Monet} D.~G.,  1989, {The Luminosity Function
  of White Dwarfs in the Local Disk and Halo}.
p.~15, \mn@doi{10.1007/3-540-51031-1\_287}

\bibitem[\protect\citeauthoryear{{Miller}, {Caiazzo}, {Heyl}, {Richer}  \&
  {Tremblay}}{{Miller} et~al.}{2022}]{2022ApJ...926L..24M}
{Miller} D.~R.,  {Caiazzo} I.,  {Heyl} J.,  {Richer} H.~B.,   {Tremblay} P.-E.,
   2022, \mn@doi [\apjl] {10.3847/2041-8213/ac50a5}, \href
  {https://ui.adsabs.harvard.edu/abs/2022ApJ...926L..24M} {926, L24}

\bibitem[\protect\citeauthoryear{{Moehler}, {Koester}, {Zoccali}, {Ferraro},
  {Heber}, {Napiwotzki}  \& {Renzini}}{{Moehler}
  et~al.}{2004}]{2004A&A...420..515M}
{Moehler} S.,  {Koester} D.,  {Zoccali} M.,  {Ferraro} F.~R.,  {Heber} U.,
  {Napiwotzki} R.,   {Renzini} A.,  2004, \mn@doi [\aap]
  {10.1051/0004-6361:20035819}, \href
  {https://ui.adsabs.harvard.edu/abs/2004A&A...420..515M} {420, 515}

\bibitem[\protect\citeauthoryear{{Mowlavi}, {Eggenberger}, {Meynet},
  {Ekstr{\"o}m}, {Georgy}, {Maeder}, {Charbonnel}  \& {Eyer}}{{Mowlavi}
  et~al.}{2012}]{2012A&A...541A..41M}
{Mowlavi} N.,  {Eggenberger} P.,  {Meynet} G.,  {Ekstr{\"o}m} S.,  {Georgy} C.,
   {Maeder} A.,  {Charbonnel} C.,   {Eyer} L.,  2012, \mn@doi [\aap]
  {10.1051/0004-6361/201117749}, \href
  {https://ui.adsabs.harvard.edu/abs/2012A&A...541A..41M} {541, A41}

\bibitem[\protect\citeauthoryear{{Munn} et~al.,}{{Munn}
  et~al.}{2017}]{2017AJ....153...10M}
{Munn} J.~A.,  et~al., 2017, \mn@doi [\aj] {10.3847/1538-3881/153/1/10}, \href
  {https://ui.adsabs.harvard.edu/abs/2017AJ....153...10M} {153, 10}

\bibitem[\protect\citeauthoryear{{Noh} \& {Scalo}}{{Noh} \&
  {Scalo}}{1990}]{1990ApJ...352..605N}
{Noh} H.-R.,  {Scalo} J.,  1990, \mn@doi [\apj] {10.1086/168562}, \href
  {https://ui.adsabs.harvard.edu/abs/1990ApJ...352..605N} {352, 605}

\bibitem[\protect\citeauthoryear{{Oswalt} \& {Smith}}{{Oswalt} \&
  {Smith}}{1995}]{1995LNP...443...24O}
{Oswalt} T.~D.,  {Smith} J.~A.,  1995, {On the Luminosity Function of White
  Dwarfs in Wide Binaries}.
p.~24, \mn@doi{10.1007/3-540-59157-5\_168}

\bibitem[\protect\citeauthoryear{{Panei}, {Althaus}, {Chen}  \& {Han}}{{Panei}
  et~al.}{2007}]{2007MNRAS.382..779P}
{Panei} J.~A.,  {Althaus} L.~G.,  {Chen} X.,   {Han} Z.,  2007, \mn@doi
  [\mnras] {10.1111/j.1365-2966.2007.12400.x}, \href
  {https://ui.adsabs.harvard.edu/abs/2007MNRAS.382..779P} {382, 779}

\bibitem[\protect\citeauthoryear{{Paxton}, {Bildsten}, {Dotter}, {Herwig},
  {Lesaffre}  \& {Timmes}}{{Paxton} et~al.}{2011}]{2011ApJS..192....3P}
{Paxton} B.,  {Bildsten} L.,  {Dotter} A.,  {Herwig} F.,  {Lesaffre} P.,
  {Timmes} F.,  2011, \mn@doi [\apjs] {10.1088/0067-0049/192/1/3}, \href
  {https://ui.adsabs.harvard.edu/abs/2011ApJS..192....3P} {192, 3}

\bibitem[\protect\citeauthoryear{{Paxton} et~al.,}{{Paxton}
  et~al.}{2013}]{2013ApJS..208....4P}
{Paxton} B.,  et~al., 2013, \mn@doi [\apjs] {10.1088/0067-0049/208/1/4}, \href
  {https://ui.adsabs.harvard.edu/abs/2013ApJS..208....4P} {208, 4}

\bibitem[\protect\citeauthoryear{{Paxton} et~al.,}{{Paxton}
  et~al.}{2015}]{2015ApJS..220...15P}
{Paxton} B.,  et~al., 2015, \mn@doi [\apjs] {10.1088/0067-0049/220/1/15}, \href
  {https://ui.adsabs.harvard.edu/abs/2015ApJS..220...15P} {220, 15}

\bibitem[\protect\citeauthoryear{{Renedo}, {Althaus}, {Miller Bertolami},
  {Romero}, {C{\'o}rsico}, {Rohrmann}  \& {Garc{\'\i}a-Berro}}{{Renedo}
  et~al.}{2010}]{2010ApJ...717..183R}
{Renedo} I.,  {Althaus} L.~G.,  {Miller Bertolami} M.~M.,  {Romero} A.~D.,
  {C{\'o}rsico} A.~H.,  {Rohrmann} R.~D.,   {Garc{\'\i}a-Berro} E.,  2010,
  \mn@doi [\apj] {10.1088/0004-637X/717/1/183}, \href
  {https://ui.adsabs.harvard.edu/abs/2010ApJ...717..183R} {717, 183}

\bibitem[\protect\citeauthoryear{{Rolland}, {Bergeron}  \&
  {Fontaine}}{{Rolland} et~al.}{2018}]{2018ApJ...857...56R}
{Rolland} B.,  {Bergeron} P.,   {Fontaine} G.,  2018, \mn@doi [\apj]
  {10.3847/1538-4357/aab713}, \href
  {https://ui.adsabs.harvard.edu/abs/2018ApJ...857...56R} {857, 56}

\bibitem[\protect\citeauthoryear{{Rowell}}{{Rowell}}{2013}]{2013MNRAS.434.1549R}
{Rowell} N.,  2013, \mn@doi [\mnras] {10.1093/mnras/stt1110}, \href
  {https://ui.adsabs.harvard.edu/abs/2013MNRAS.434.1549R} {434, 1549}

\bibitem[\protect\citeauthoryear{{Rowell} \& {Hambly}}{{Rowell} \&
  {Hambly}}{2011}]{2011MNRAS.417...93R}
{Rowell} N.,  {Hambly} N.~C.,  2011, \mn@doi [\mnras]
  {10.1111/j.1365-2966.2011.18976.x}, \href
  {https://ui.adsabs.harvard.edu/abs/2011MNRAS.417...93R} {417, 93}

\bibitem[\protect\citeauthoryear{{Salaris}, {Dom{\'\i}nguez},
  {Garc{\'\i}a-Berro}, {Hernanz}, {Isern}  \& {Mochkovitch}}{{Salaris}
  et~al.}{1997}]{1997ApJ...486..413S}
{Salaris} M.,  {Dom{\'\i}nguez} I.,  {Garc{\'\i}a-Berro} E.,  {Hernanz} M.,
  {Isern} J.,   {Mochkovitch} R.,  1997, \mn@doi [\apj] {10.1086/304483}, \href
  {https://ui.adsabs.harvard.edu/abs/1997ApJ...486..413S} {486, 413}

\bibitem[\protect\citeauthoryear{{Salaris}, {Serenelli}, {Weiss}  \& {Miller
  Bertolami}}{{Salaris} et~al.}{2009}]{2009ApJ...692.1013S}
{Salaris} M.,  {Serenelli} A.,  {Weiss} A.,   {Miller Bertolami} M.,  2009,
  \mn@doi [\apj] {10.1088/0004-637X/692/2/1013}, \href
  {https://ui.adsabs.harvard.edu/abs/2009ApJ...692.1013S} {692, 1013}

\bibitem[\protect\citeauthoryear{{Salaris}, {Cassisi}, {Pietrinferni},
  {Kowalski}  \& {Isern}}{{Salaris} et~al.}{2010}]{2010ApJ...716.1241S}
{Salaris} M.,  {Cassisi} S.,  {Pietrinferni} A.,  {Kowalski} P.~M.,   {Isern}
  J.,  2010, \mn@doi [\apj] {10.1088/0004-637X/716/2/1241}, \href
  {https://ui.adsabs.harvard.edu/abs/2010ApJ...716.1241S} {716, 1241}

\bibitem[\protect\citeauthoryear{{Schlafly} \& {Finkbeiner}}{{Schlafly} \&
  {Finkbeiner}}{2011}]{2011ApJ...737..103S}
{Schlafly} E.~F.,  {Finkbeiner} D.~P.,  2011, \mn@doi [\apj]
  {10.1088/0004-637X/737/2/103}, \href
  {https://ui.adsabs.harvard.edu/abs/2011ApJ...737..103S} {737, 103}

\bibitem[\protect\citeauthoryear{{Schmidt}}{{Schmidt}}{1959}]{1959ApJ...129..243S}
{Schmidt} M.,  1959, \mn@doi [\apj] {10.1086/146614}, \href
  {https://ui.adsabs.harvard.edu/abs/1959ApJ...129..243S} {129, 243}

\bibitem[\protect\citeauthoryear{{Shaviv} \& {Kovetz}}{{Shaviv} \&
  {Kovetz}}{1976}]{1976A&A....51..383S}
{Shaviv} G.,  {Kovetz} A.,  1976, \aap, \href
  {https://ui.adsabs.harvard.edu/abs/1976A&A....51..383S} {51, 383}

\bibitem[\protect\citeauthoryear{{Temmink}, {Toonen}, {Zapartas}, {Justham}  \&
  {G{\"a}nsicke}}{{Temmink} et~al.}{2020}]{2020A&A...636A..31T}
{Temmink} K.~D.,  {Toonen} S.,  {Zapartas} E.,  {Justham} S.,   {G{\"a}nsicke}
  B.~T.,  2020, \mn@doi [\aap] {10.1051/0004-6361/201936889}, \href
  {https://ui.adsabs.harvard.edu/abs/2020A&A...636A..31T} {636, A31}

\bibitem[\protect\citeauthoryear{{Tremblay} \& {Bergeron}}{{Tremblay} \&
  {Bergeron}}{2009}]{2009ApJ...696.1755T}
{Tremblay} P.~E.,  {Bergeron} P.,  2009, \mn@doi [\apj]
  {10.1088/0004-637X/696/2/1755}, \href
  {https://ui.adsabs.harvard.edu/abs/2009ApJ...696.1755T} {696, 1755}

\bibitem[\protect\citeauthoryear{{Tremblay}, {Bergeron}  \&
  {Gianninas}}{{Tremblay} et~al.}{2011}]{2011ApJ...730..128T}
{Tremblay} P.~E.,  {Bergeron} P.,   {Gianninas} A.,  2011, \mn@doi [\apj]
  {10.1088/0004-637X/730/2/128}, \href
  {https://ui.adsabs.harvard.edu/abs/2011ApJ...730..128T} {730, 128}

\bibitem[\protect\citeauthoryear{{Virtanen} et~al.,}{{Virtanen}
  et~al.}{2020}]{2020NatMe..17..261V}
{Virtanen} P.,  et~al., 2020, \mn@doi [Nature Methods]
  {10.1038/s41592-019-0686-2}, \href
  {https://ui.adsabs.harvard.edu/abs/2020NatMe..17..261V} {17, 261}

\bibitem[\protect\citeauthoryear{{Williams}, {Bolte}  \& {Koester}}{{Williams}
  et~al.}{2009}]{2009ApJ...693..355W}
{Williams} K.~A.,  {Bolte} M.,   {Koester} D.,  2009, \mn@doi [\apj]
  {10.1088/0004-637X/693/1/355}, \href
  {https://ui.adsabs.harvard.edu/abs/2009ApJ...693..355W} {693, 355}

\bibitem[\protect\citeauthoryear{{Winget}, {Hansen}, {Liebert}, {van Horn},
  {Fontaine}, {Nather}, {Kepler}  \& {Lamb}}{{Winget}
  et~al.}{1987}]{1987ApJ...315L..77W}
{Winget} D.~E.,  {Hansen} C.~J.,  {Liebert} J.,  {van Horn} H.~M.,  {Fontaine}
  G.,  {Nather} R.~E.,  {Kepler} S.~O.,   {Lamb} D.~Q.,  1987, \mn@doi [\apjl]
  {10.1086/184864}, \href
  {https://ui.adsabs.harvard.edu/abs/1987ApJ...315L..77W} {315, L77}

\bibitem[\protect\citeauthoryear{{Wood}}{{Wood}}{1992}]{1992ApJ...386..539W}
{Wood} M.~A.,  1992, \mn@doi [\apj] {10.1086/171038}, \href
  {https://ui.adsabs.harvard.edu/abs/1992ApJ...386..539W} {386, 539}

\bibitem[\protect\citeauthoryear{{Zhao}, {Oswalt}, {Willson}, {Wang}  \&
  {Zhao}}{{Zhao} et~al.}{2012}]{2012ApJ...746..144Z}
{Zhao} J.~K.,  {Oswalt} T.~D.,  {Willson} L.~A.,  {Wang} Q.,   {Zhao} G.,
  2012, \mn@doi [\apj] {10.1088/0004-637X/746/2/144}, \href
  {https://ui.adsabs.harvard.edu/abs/2012ApJ...746..144Z} {746, 144}

\bibitem[\protect\citeauthoryear{{Zuckerman}, {Xu}, {Klein}  \&
  {Jura}}{{Zuckerman} et~al.}{2013}]{2013ApJ...770..140Z}
{Zuckerman} B.,  {Xu} S.,  {Klein} B.,   {Jura} M.,  2013, \mn@doi [\apj]
  {10.1088/0004-637X/770/2/140}, \href
  {https://ui.adsabs.harvard.edu/abs/2013ApJ...770..140Z} {770, 140}

\makeatother
\end{thebibliography}






\bsp	
\label{lastpage}
\end{document}